\documentclass[useAMS,usenatbib,usegraphicx]{mn2e}  
\usepackage{longtable}
\usepackage{lastpage}
\usepackage{color}
\newcommand{\ion}[2]{#1\,{\mdseries\textsc{#2}}}
\newcommand{\nickel}{$^{56}$Ni}
\newcommand{\vphot}{$v_\textrm{ph}$}
\newcommand{\kms}{\,km\,s$^{-1}$}

\definecolor{red}{rgb}{1.00,0.00,0.00}
\definecolor{blue}{rgb}{0.00, 0.00, 1.00}

\bibpunct{(}{)}{;}{a}{}{,}

\title[Studying the Diversity of SNe Ia in the UV]{Studying the Diversity of Type Ia Supernovae in the Ultraviolet: Comparing Models with Observations}

   \author[E.S. Walker et al.]
{
\parbox{\textwidth}
{E.S.~Walker $^{1}$\thanks{E-mail address emma.walker@sns.it}, S. Hachinger $^2$, P.A. Mazzali $^{2,3}$, R.S.~Ellis $^{4}$, M.~Sullivan $^{5}$, \mbox{A.~Gal-Yam $^{6}$}, D.A.~Howell $^{7,8}$}
\vspace{0.4cm}\\
\parbox{\textwidth}
{$^1$  Scuola Normale Superiore di Pisa, Piazza dei Cavalieri 7, 56126 Pisa, Italy\\
$^2$   INAF--Osservatorio Astronomico, vicolo dell'Osservatorio, 5, I-35122 Padova, Italy\\
$^3$   Max-Planck Institut f\"ur Astrophysik, Karl-Schwarzschildstr. 1, D-85748 Garching, Germany \\
$^4$ California Institute of Technology, East California Boulevard, Pasadena, CA 91125 USA \\
$^5$ Oxford Astrophysics, Denys Wilkinson Building, Keble Road, Oxford, OX1 3RH, UK\\
$^6$ Department of Particle Physics and Astrophysics, Faculty of Physics, The Weizmann Institute of Science, Rehovot 76100, Israel\\
$^{7}$ Las Cumbres Observatory Global Telescope Network, 6740 Cortona Dr.,
Suite 102, Goleta, CA 93117, USA\\
$^{8}$Department of Physics, University of California, Santa Barbara,
Broida Hall, Mail Code 9530, Santa Barbara, CA 93106-9530, USA\\
}}
\begin{document}

\date{Released 2002 Xxxxx XX}

\pagerange{\pageref{firstpage}--\pageref{LastPage}} \pubyear{2002}

\maketitle

\label{firstpage}

\begin{abstract}

In the ultraviolet (UV), Type Ia supernovae (SNe Ia) show a much larger diversity
in their properties than in the optical.   Using a stationary Monte-Carlo
radiative transfer code, a grid of spectra at maximum light was created varying
bolometric luminosity and the amount of metals in the outer layers of the SN
ejecta.  This model grid is then compared to a sample of high-redshift SNe Ia
in order to test whether the observed diversities can be explained by luminosity
and metallicity changes alone.  The dispersion in broadband UV flux and colours
at approximately constant optical spectrum can be readily matched by the model
grid. In particular, the UV1-b colour is found to be a good tracer of metal
content of the outer ejecta, which may in turn reflect on the metallicity of the
SN progenitor. The models are less successful in reproducing other observed
trends, such as the wavelengths of key UV features, which are dominated by
reverse fluorescence photons from the optical, or intermediate band photometric
indices. This can be explained in terms of the greater sensitivity of these
detailed observables to modest changes in the relative abundances. Specifically,
no single element is responsible for the observed trends. Due to their
complex origin, these trends do not appear to be good indicators of either
luminosity or metallicity. 

\end{abstract}


 \section{Introduction}\label{sec-intro}
 
Type Ia supernovae (SNe Ia) are some of the most important tools for current
cosmological studies.  Following the discovery that their peak magnitudes could
be standardised \citep{P93}, their use enabled the discovery that the universe
is accelerating \citep{R98,P99}.  More recent supernova observations, combined
with other constraints from the cosmic microwave background and baryon acoustic
oscillations have established that we live in a flat universe with a matter
content of $\Omega_M \approx 0.27$
\citep{Astier06,WoodVasey:2007ky,Kessler:2009et,Sullivan:2011bu} and the
remaining 73\%\ made up of dark energy, the nature of which is currently
unknown.

SNe Ia studies have measured that $w$, the dark energy equation of state
parameter is consistent with $w = -1$ to 6.5\% with some studies at high-z even
beginning to place constraints on whether $w = w(z)$ \citep{Riess07}.  In the
future, we will have to observe at higher redshifts in order to find supernovae
from younger times in the universe to improve on these dynamical measurements. 
As such our surveys will either have to switch to the IR or probe the rest-frame
UV, a region of the spectrum that has been less extensively explored in the
local population.

The UV spectra of SNe Ia have long been thought to probe the region where
metallicity effects would be important \citep{Hoflich:1998eg,Lentz:2000ff} due
to the vast number of metal line transitions in this region.  Many of the
photons in this region are absorbed, mostly by iron group elements, in what is
referred to as metal line-blanketing; however, the effect that metallicity has
on the level of the continuum flux in this region is debated
\citep{Sauer:2008gm}.

One example of the effect that progenitor metallicity may have is that in a
higher metallicity progenitor, the production of neutron-rich isotopes such as
$^{54}$Fe and $^{58}$Ni is favoured compared to \nickel\ \citep{Iwamoto:1999}.
This will be reflected not only in the spectra, but also to some degree in the
broad-band light curves. A SN Ia in the local universe, where metallicity is
high, will on average have a lower luminosity than a high-redshift SN Ia due to
the different \nickel\ content and hence a different light curve stretch, as
shown in \citet{Timmes:2003bp,Howell:2007fl,Howell:2009fm}.  The abundance ratio
of stable iron-group elements to radioactive \nickel\ has been proposed as an
additional parameter for the standardisation of SN Ia light curves
\citep{Mazzali:2006cy}.

Until recently, our understanding of the role of progenitor metallicity in
SNe\,Ia data was limited by the paucity of observed UV spectra. Recognising
this, \citet[][hereafter, E08]{Ellis08} secured high quality Keck  spectra for
36 intermediate redshift ($z\simeq$0.5) SNe\,Ia at maximum light drawn from the
Supernova Legacy Survey (SNLS); these optical spectra appropriately probe the
rest-frame UV. \defcitealias{Ellis08}{E08} \citetalias{Ellis08} noted a
significant diversity in their UV spectra which could not be attributed to dust.
Importantly, they found the variations in their UV data, as characterised by
colours derived directly from their rest-frame spectra, did  not correlate with
the light-curve stretch. They also showed that the wavelengths of specific UV
features showed phase-dependent shifts. The dispersion in their UV colours was
claimed to be larger than could be accounted for metal-dependent models
available at the time \citep{Lentz:2000ff} thus opening the possibility of an
additional explanation for the diversity.

Recently, Hubble Space Telescope (HST) and Swift observations have begun to
explore the rest-frame UV of local SNe Ia.  \citet{Foley:2008fx} used archival
HST and International Ultraviolet Explorer (IUE) data to show that a particular
ratio of UV flux correlates strongly with absolute V-band magnitude for 6
objects with spectra near maximum light: brighter supernovae have lower values
of the ratio.  This is a different result from that claimed in
\citetalias{Ellis08} which saw no correlation with supernova brightness.

In a more recent study using 21 intermediate redshift SNe Ia ($z\simeq$0.25), \citet{Foley:2012ej} repeat the UV flux ratio analysis and find a different relation between absolute V-band magnitude and the ratio value.  In this case, brighter supernovae still show lower values of the ratio value, but the slope of the relation is very different.
\citet{Foley:2008fx} explored the use of the UV ratio as a luminosity indicator
for light curve standardisation with some degree of success.  The potential link
between UV properties and intrinsic luminosity would have important implications
for future cosmological studies.

Swift data have been used by \citet{Brown:2010cd} and \citet{Milne:2010em} to
obtain an overview of the spectral behaviour in the UV. In the near-UV filters
(2600--3300\AA\ and 3000--4000\AA), the normal sub-class of SNe Ia shows a high
degree of homogeneity, while the subluminous and the peculiar SN\,2002cx-like
groups show large differences. Absolute magnitudes at maximum brightness are
correlated with the optical decay rate and show a scatter similar in size to
that obtained with optical data.  However, in the mid-UV (2000--2400\AA) the
scatter is much larger ($\sim 1$\,mag), indicating possible metallicity-driven
effects in this part of the spectrum. 

Recent HST observations of 12 Hubble-flow SNe Ia at maximum light by
\citet{Cooke:2011jm} show that the dispersion from a mean spectrum increases as
wavelength decreases, and is largest in the UV region of the spectrum.  They
attribute this to the larger number of metal absorption lines in the UV compared
to the optical.   The same increase in dispersion is seen at higher redshift to
the same degree \citepalias{Ellis08} so they conclude that this must be an
intrinsic feature of the supernova and not due to evolutionary effects.  A
larger study of UV spectra at maximum from HST is underway (Maguire et al.~2012,
in press).  A large degree of diversity is also seen in the UV photometry and spectra of four supernovae discussed in \citet{Wang:2012cn}.  The paper concludes that more detailed modelling of supernovae in the UV is required.

Optical studies have shown that supernova properties depend on the properties of
the host galaxy.  \citet{Hamuy:2000il} first showed that the higher mass
galaxies preferentially host dimmer supernovae compared to brighter supernovae
which were associated with younger stellar populations in late-type galaxies.

\citet{Sullivan:2010ez} examined SNe Ia subdividing their sample by host properties.  They found that in more massive galaxies, or in those with a lower specific star-formation rate, the SNe Ia were on average $\simeq 0.08$\,mag brighter than that of SNe Ia in other galaxies after correction for light-curve stretch and colour. \citet{Sullivan:2010ez} suggested that the difference they observe may be due to the metallicity of the host galaxy as more massive galaxies tend to be more metal-rich; however this appears at odds with the results of \citet {Timmes:2003bp} and \citet{Mazzali:2006cy}.

This study is thus motivated by the need to reconcile the conflicting deductions
regarding the observed diversity in the UV spectra derived from earlier work  We
exploit a wide range of models parameterised by both bolometric luminosity and
metallicity to see if we can explain the observations with these two variables
alone.  The model dataset is presented in Section \ref{sec:model-dataset} and
the optical data sample which we use for comparison is described in Section
\ref{sec:obs-data}.  In Section \ref{sec:analysis} we compare measurements of
various UV properties of the model and data samples.  Our results are then
discussed in Section \ref{sec-discussion} and a summary of our conclusions
presented in Section \ref{sec:conclusions}.

\section{Datasets}\label{sec:dataset}

\subsection{Model Dataset}\label{sec:model-dataset}

We calculate model spectra in order to study the influence of metallicity and
luminosity on the UV spectra of ``normal'' SNe Ia around $B$-band maximum. The
respective radiative transfer models are based on those in \citet{Sauer:2008gm}.
A two-parameter grid of models is set up by changing the bolometric luminosity
of the models and the metallicity in the outer ejecta to study the effect of these parameters on the UV.

\subsubsection{Code and Model Input}

To calculate the synthetic spectra, we ran a stationary Monte-Carlo radiative
transfer code, which has successfully been used to model photospheric spectra of
numerous SNe Ia in spherical symmetry
\citep{Mazzali:1993vl,Lucy:1999wl,Mazzali:2000va,Stehle:2005ff}, including rare
UV observations \citep{Sauer:2008gm}.

The version of the code used here simulates a SN atmosphere above a lower
boundary (``photosphere''), and takes as input data the location of the
photosphere, $r_\mathrm{ph}$ or $v_\mathrm{ph}$; the time passed from explosion
onset, $t$; an abundance stratification; and a density profile. Here, we use the
density profile of the SN Ia delayed detonation model WDD3 \citep{Iwamoto:1999}.
The profile is scaled by the code, assuming homologous expansion to time $t$,
i.~e.~for each mass element $r= vt$ (where $r$ is the distance from the centre
of mass and $v$ is the velocity imparted at explosion).  Thus either $r$ or $v$
can be used as spatial coordinates.  Apart from the parameters mentioned, the
code allows the user to set a bolometric luminosity, $L_{bol}$, for the final
output spectrum.

From the photosphere, which is located at an adjustable $v_\mathrm{ph}$, energy
packets of continuous black body radiation ($I_{\nu}^{+} =
B_{\nu}T_\mathrm{ph}$) are assumed to be emitted into the atmosphere. The
simulated photon packets undergo Thomson scattering and interactions with atomic
absorption/emission lines, which are treated in the Sobolev approximation. A
downward branching scheme ensures a good approximation to the bound-bound
emissivity. Radiative equilibrium is enforced by the construction of the
Monte-Carlo simulation \citep{Lucy:1999wl}. Bound-free processes are not
simulated, as lines together with Thomson scattering dominate the opacity in SNe
Ia \citep{Pauldrach:1996ts,Sauer:2006gj}.

The excitation and ionisation state of the matter are calculated from the
radiation field statistics using a modified nebular approximation
\citep{Mazzali:1993vl,Mazzali:2000va}. In this approximation, the gas state in
each radial grid cell is mostly determined from a radiation temperature $T_{R}$
and a dilution factor $W$. $T_{R}$ corresponds to the mean frequency of the
radiation field and $W$ parametrises its energy density (for given $T_{R}$). We
iterate the state of the gas and the radiation field in turn until the $T_{R}$
values within the atmosphere are converged to the per cent level. Within these
iterations, $T_\mathrm{ph}$ is automatically adjusted so as to match the given
luminosity $L_{bol}$. This adjustment compensates for the reabsorption of
radiation which occurs when packets re-enter the photosphere via
back-scattering. After the iterations have converged, the final spectrum is
calculated by formal integration of the transfer equation \citep{Lucy:1999wl},
using a source function derived from the Monte Carlo statistics.

\subsubsection{Model Grid}

\begin{figure}
\centering
\includegraphics[width=\columnwidth]{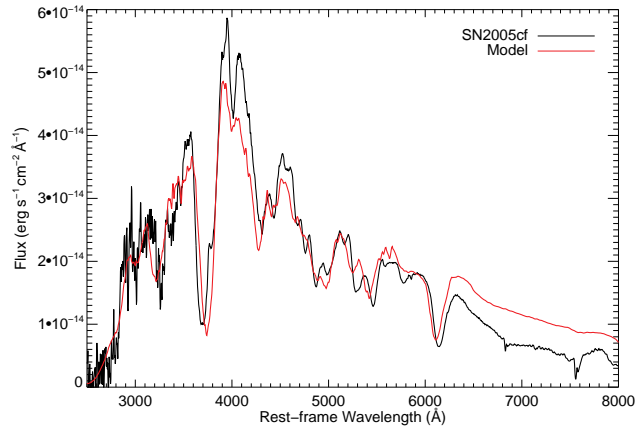}
\caption{Synthetic spectrum from our ejecta model (red line) compared to an observed UV-optical spectrum (black line) taken 0.9\,days before $B$ maximum \protect\citep{Bufano:2009gp,Garavini:2007bu}.}
\label{fig:05cf-spectrum}
\end{figure}

All synthetic spectra in this paper are derived from a model developed for
SN\,2005cf, for which UV/optical data near maximum is available
\citep{Bufano:2009gp,Garavini:2007bu,Wang:2009bf}.  SN\,2005cf was a bright supernova ($M_B = -19.39$ \citep{Pastorello:2007er}) with a light-curve width $\Delta m_{15} = 1.10$ \citep{Pastorello:2007er,Wang:2009bf} which corresponds to a light-curve stretch of $s = 0.93$ using the relation in \citet{sifto}.  Since SN\,2005cf was a
bright SN and produced a large amount of $^{56}$Ni ($\approx 0.7M_{\odot}$) we chose to use as the density structure of
model WDD3 from \citet{Iwamoto:1999}.  The bolometric luminosity of
this model is $\log(L_{bol}/L_{\odot}) = 9.6$. 

We have used an abundance tomography approach to model the UVOIR spectrum of
SN\,2005cf 0.9 days before maximum. The method of successively constraining the
abundances in deeper and deeper layers using a temporal sequence of spectra was
first introduced by \citet{Stehle:2005ff}.  In order to constrain properly the
highest velocity material in the outermost ejecta before trying to model a
spectrum at maximum, we started with a spectrum obtained $-7.8$\,days before 
$B$-band maximum.  The maximum brightness spectrum has a phase of $-0.9$\,days
relative to $B$-band maximum.  We assume a B-band rise-time of $19.5$\,days
\citep{Conley:2006cx} and use $t=18.6$ days.  In order to optimise the fits to
the data, two additional zones were introduced on top of those at
maximum and pre-maximum velocities, as in \cite{Sauer:2008gm}.  This is not
unexpected as strong high-velocity features have been noted in SN\,2005cf
\citep{Garavini:2007bu}.  We optimised the match to the UV
spectrum ($\lambda < 4000$\,\AA) while maintaining the best possible fit in the optical.

This model is reproduced in Figure \ref{fig:05cf-spectrum}.  A summary of the
chemical composition in each zone is given in Table \ref{tab:05cf_massfrac} where IME
stands for intermediate mass elements (those with atomic number 9 -- 20) and IGE for iron-group elements (atomic number greater than 20).

From Table \ref{fig:05cf-spectrum} we see that the outer two layers are
dominated by unburnt carbon and oxygen with very little IME and almost no IGE. 
In the pre-maximum layer, there is still some C/O material remaining and
slightly more IGE material, but the shell is dominated by material that has been
burnt to IME.  The shell at the photospheric velocity at maximum is still
dominated by IME, but now a significant fraction of the shell is made up of IGE.

\begin{table}
\caption{A summary of the major element groups in each shell of the model for SN\,2005cf.  X is the mass fraction of C/O, intermediate mass elements (IME) and iron group elements (IGE).}
\label{tab:05cf_massfrac}
\centering
\begin{tabular}{l  c c c c }
\hline
Zone & \vphot & $v_{\textrm{pre-max}}$ & $v_{\textrm{out,1}}$ & $v_{\textrm{out,2}}$\\
\hline
Velocity & 10750 & 13100 & 16000 & 19500 \\
X(C/O) & 0.3 & 4.0 & 70 & 92\\
X(IME) & 63 & 92 & 18 & 8.2\\
X(IGE) & 37 & 3.9 & 3.2 & 0.2\\
\hline
\end{tabular}
\end{table}

\begin{figure}
\centering
\includegraphics[width =\columnwidth]{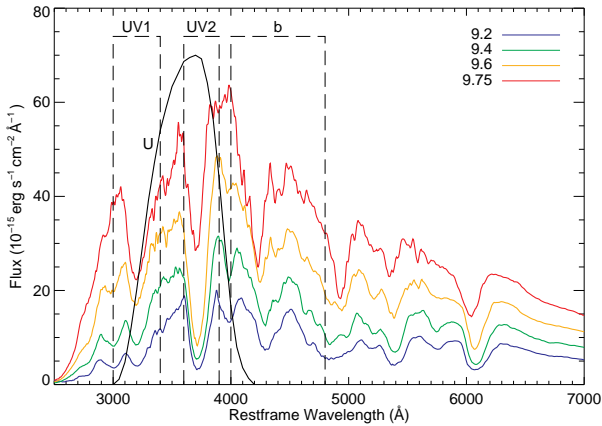}
\caption{The $\eta = 1$ spectra for $\log(L_{bol}/L_{\odot})=\textrm{9.2, 9.4, 9.6, 9.75}$.  Also marked are the three top-hat filters introduced in \citetalias{Ellis08}, UV1, UV2 and b (dashed lines) and the Bessel U filter (solid black line).}
\label{fig:model_lum}
\end{figure}

We used this technique to create models for supernovae with different bolometric
luminosities.  The range of luminosities was chosen to reflect the spread of observed SN\,Ia properties.  We also used
different underlying density profiles to reflect the different energies and
\nickel\ production at the bolometric luminosities $\log(L_{bol}/L_{\odot}) =
9.2,9.4,9.6,9.75$. The velocities of the \vphot\ and $v_\textrm{pre-max}$
shells move depending on luminosity (Table 2), resulting in the masses of individual elements being scaled
for the whole model. However, within each shell the relative
abundances of the SN\,2005cf model are preserved as described.  In order to
ensure that we produce realistic models, we compare the model output spectrum to
observed SNe Ia to ensure we are producing spectra which match the continuum
levels in the UV while maintaining normal optical spectra i.~e.~ are not members of the over-luminous of under-luminous subclasses.  This process is
summarised in Table \ref{tab:bol_lum_params} and the spectra are displayed in
Figure \ref{fig:model_lum}

\begin{table}
\caption{An overview of the model parameters for the models $\log(L_{bol}/\L_{\odot})=\textrm{9.2, 9.4, 9.6, 9.75}$ with $\eta = 1$.  The W7 model is from \citet{N84} and the WDD1 and WDD3 models are from \citep{Iwamoto:1999}.  The given velocities are in \kms.}
\label{tab:bol_lum_params}
\centering
\begin{tabular}{lcccc}
\hline
& 9.2 & 9.4 & 9.6 & 9.75\\
\hline
Density Structure & W7 & WDD1 & WDD3 & WDD3\\
\vphot & 6783 & 8539 & 10750 & 11250\\
$v_{\textrm{pre-max}}$ & 8366 & 10406 & 13100 & 13710\\
$v_{\textrm{out,1}}$ & \multicolumn{4}{c}{16000}\\
$v_{\textrm{out,2}}$ & \multicolumn{4}{c}{19500}\\
\hline
\end{tabular}
\end{table}

\begin{figure}
\centering
\includegraphics[width =\columnwidth]{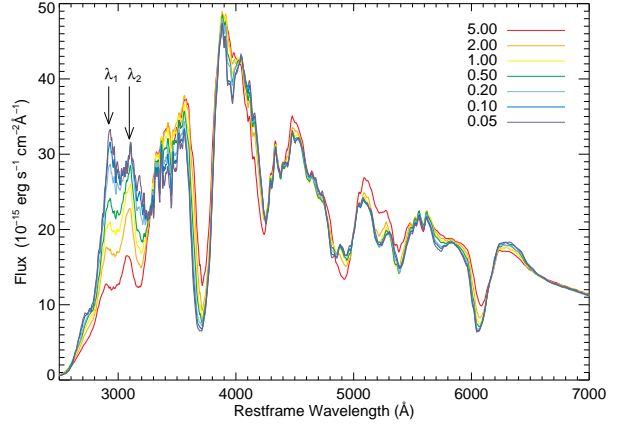}
\caption{The $\eta = 5,2,1,0.5,0.2,0.1,0.05$ spectra for $\log(L_{bol}/L_{\odot})=\textrm{9.6}$.  The features $\lambda_1$ and $\lambda_2$ are also marked.}
\label{fig:model_met}
\end{figure}

We then used these 4 luminosity bins to create a set of models with varying
metal contents.  We do this by scaling the metallicity in the pre-maximum and
two outer shells with respect to the best-fitting model: the photospheric shell
remains unchanged.  To generate the sequences, we multiplied the abundances of all
the elements with atomic number $Z>20$, i.~e. heavier than calcium, by a factor
$\eta$, which is allowed to take the values $\eta = 0.05, 0.1, 0.2, 0.5, 1, 2, 5$.  The mass fraction $X$ of element $E$ thus becomes $X(E) = \eta X(E)_0$
where $X(E)_0$ is the mass fraction of the element in the \emph{SN\,2005cf best-fit
model} at the expense of unburnt C/O. This
provides us with a grid of $4\times 7$ models for our analysis, but we have
excluded the model where $\log(L_{bol}/L_{\odot})=\textrm{9.75}$ and $\eta =
0.05$ as the optical spectrum did not look normal, leaving 27 models for
analysis.  A metallicity sequence is shown in Figure \ref{fig:model_met} for
$\log(L_{bol}/L_{\odot})=\textrm{9.6}$.  For all values of $\eta$, the optical
spectra appear normal with relatively little dispersion. However, in the UV the
dispersion between the models increases dramatically and diversity is also
seen in the shapes and positions of features.  This reflects the fact that 
metal line-blanketing effect is stronger in the UV.

One caveat with our models is that in the red and infrared the fits to data are
less good.  This is because of the crude assumption of a blackbody at the
photosphere.  The flux inside an SN Ia is non-thermal even in the inner layers
\citep[see][]{Sauer:2006gj}.  Flux redistribution within the inner parts of the
simulated atmosphere leads to a sufficiently accurate radiation field in the
atmosphere in the ultraviolet and blue regions of the spectrum, but in the red
and infrared some flux excess usually remains in the synthetic spectra with
respect to observations.  This can be seen in Figure \ref{fig:05cf-spectrum}
where the model flux is higher than the observed flux from $\approx6300$\,\AA\
onwards.  This means that our estimates of $L_{Bol}$ may be somewhat larger than
the real value when the red and IR are overestimated. Therefore, in order to
compare models and data we extract $L_B$ from both.

\subsection{Observational Data}\label{sec:obs-data}

The observational data for this study are taken from those presented in
\citetalias{Ellis08}. These SNe were discovered as part of the Supernova Legacy
Survey \citep[SNLS,][]{Sullivan:2011bu}, a real-time  Type Ia supernova search
based at the Canada-France-Hawaii Telescope (CFHT) which used $g'r'i'z'$-band
observations to identify high-z SNe Ia.  For more details on the real-time SNLS
target selection pipe-line see \citet{Perrett:2010ku}.

\citetalias{Ellis08} observed a selection of SNLS supernovae on the Keck I Telescope,
using LRIS \citep{Oke:1995ku} to obtain a high signal-to-noise ratio in the
rest-frame UV. Using host photometry obtained from the deep stack CFHT images in
the $u^*g'r'i'z'$-bands, a best-fitting template galaxy spectrum was used to
remove contamination from host galaxy light in the SN spectrum.  For more
details on this see \citetalias{Ellis08}, or \citet{2011MNRAS.410.1262W} which gives
a detailed explanation of the application of this method to SNLS data obtained
at the Gemini and VLT telescopes.

It is important to note that while the sample of supernovae used in
\citetalias{Ellis08} as a whole was representative of the supernova population,
the sub-sample of these spectra we use here are not because we apply a cut for
rest-frame phase.  In this study, in order to make a realistic comparison of
models to data, we included only supernovae with a rest-frame phase of
$\pm3$\,days from $B$-band maximum.  Additionally, we only considered "normal"
SNe Ia with good lightcurve coverage so the $B$-band maximum magnitude could be
calculated, and with a spectrum reaching a minimum rest-frame wavelength of $\leq
2700$\AA.  These cuts leave 9 objects.

Within our sub-sample the mean light-curve stretch is $\overline{s} = 1.08$,
where 1.0 represents the fiducial "normal" SN Ia lightcurve.  In fact, within our
sub-sample, all but one of the objects have a stretch value $>1$.  The mean
colour of our subsample is $\overline{c} = -0.027$.  The light-curve fitting was
carried out using the SiFTO fitter \citep{sifto}.  SiFTO also fits a $B$-band
magnitude at maximum.  To obtain $L_B$ we convert magnitude to flux and then to
luminosity assuming a flat cosmology with $\Omega_M = 0.269$
\citep{Sullivan:2011bu} and $H_0 = 70$\kms\,Mpc$^{-1}$.

\begin{figure}
 \centering
\includegraphics[width =\columnwidth]{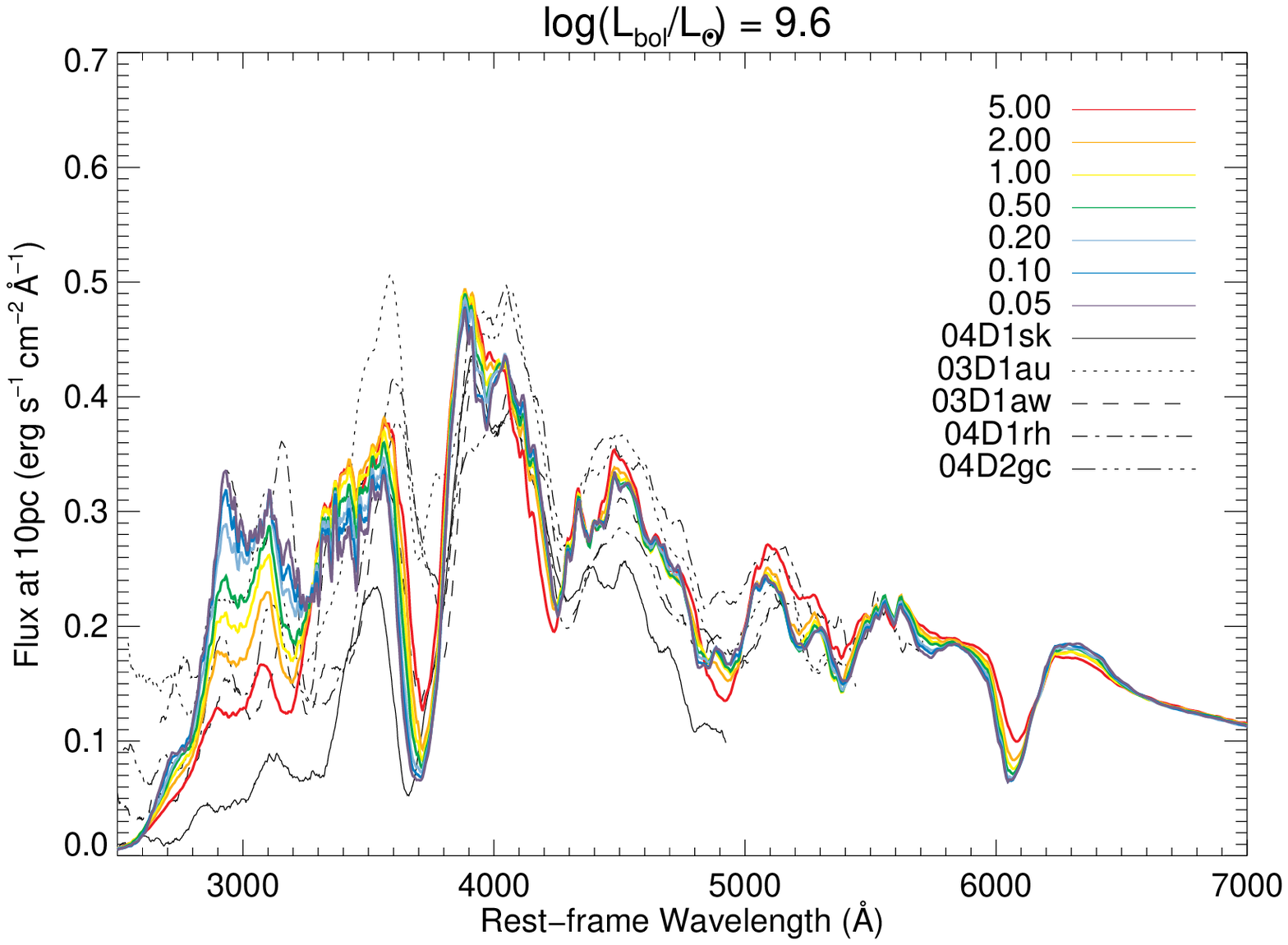}
\includegraphics[width =\columnwidth]{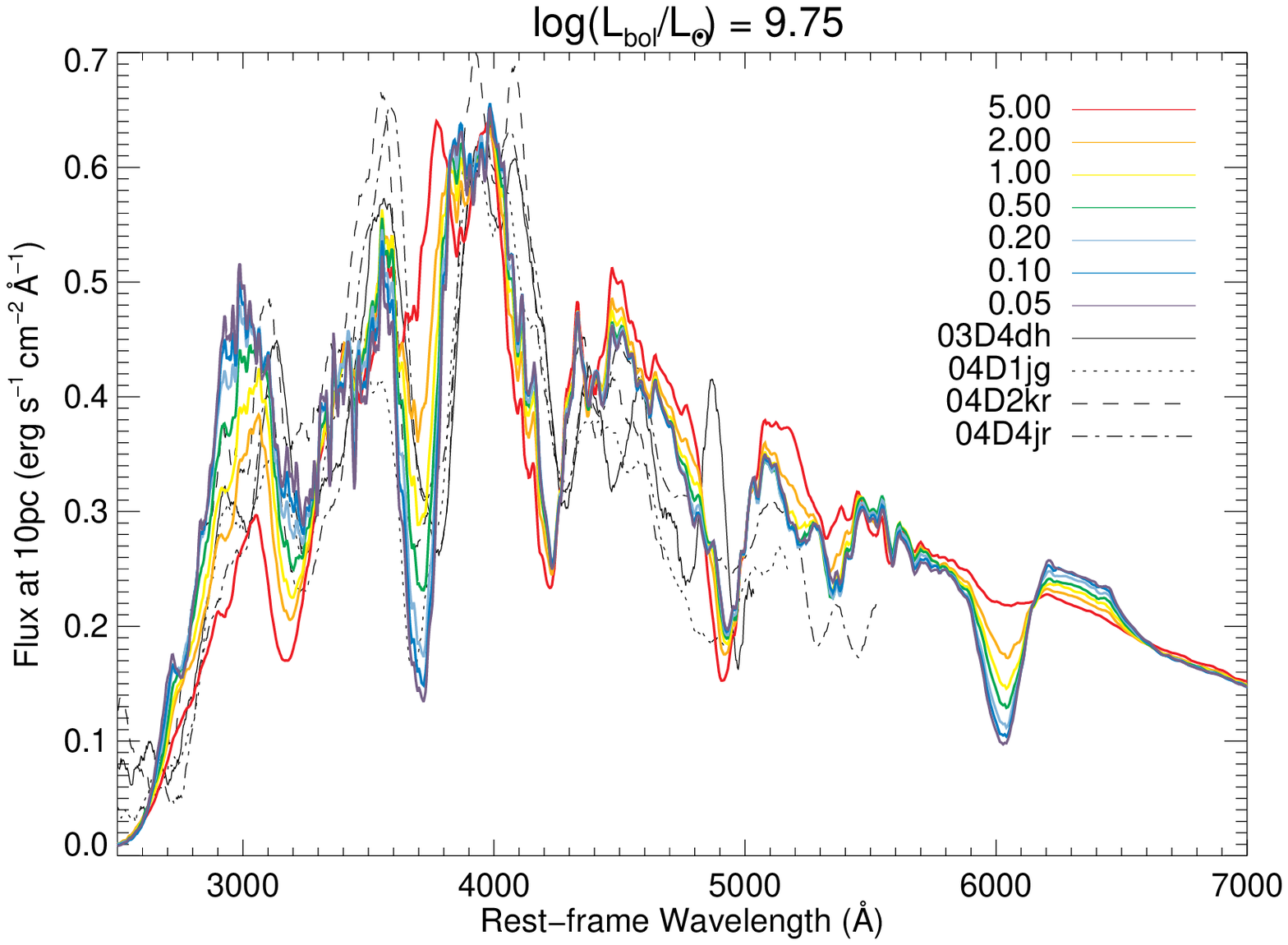}
\caption{For all values of $\eta$, the models with $\log(L_{bol}/L_{\odot})=\textrm{9.6}$ (top) and $\log(L_{bol}/L_{\odot})=\textrm{9.75}$ (bottom) are compared to observed spectra from \citetalias{Ellis08}.}
\label{fig:model_data_comp}
\end{figure}

In Figure \ref{fig:model_data_comp} we plot the various models at
$\log(L_{bol}/L_{\odot})=\textrm{9.6, 9.75}$ and compare them to our
observational data assuming all supernovae are placed at 10\,pc.  The variations
in metallicity appear to match the variations in the spectra of the
\citetalias{Ellis08} sample.  The spectrum lying below the models and the other
observed SNe Ia with $\log(L_{bol}/L_{\odot})=\textrm{9.6}$ is 04D1sk.  This
appears to have a low UV flux compared to the optical.  As such, we expect this
to be an outlier in some of our UV colour analysis (Section
\ref{sec:uvcolours}).  None of the observed supernovae have luminosities which
would correspond to the $\log(L_{bol}/L_{\odot})=\textrm{9.2,9.4}$ models; however  SNe Ia are observed to have luminosities in this range (see Figure \ref{fig:bmag} for example).

\section{Analysis}\label{sec:analysis}

We now have a set of models and observed spectra which can be analysed in
identical ways.  As the $B$-band magnitudes of the observed supernovae are
well-known from their use in cosmology, we revert to using the $B$-band
luminosity $L_B$ instead of bolometric luminosity as this quantity is
measureable for both the models and observed supernovae.  We can therefore make
direct comparisons between the observed and model data for the UV diagnostics
described in this section.  It is possible to see from later plots that at
constant $L_{bol}$, $L_B$ does not vary linearly with $\eta$.

\subsection{UV Colours}\label{sec:uvcolours}

\begin{figure*}
\centering
\includegraphics[width=\textwidth]{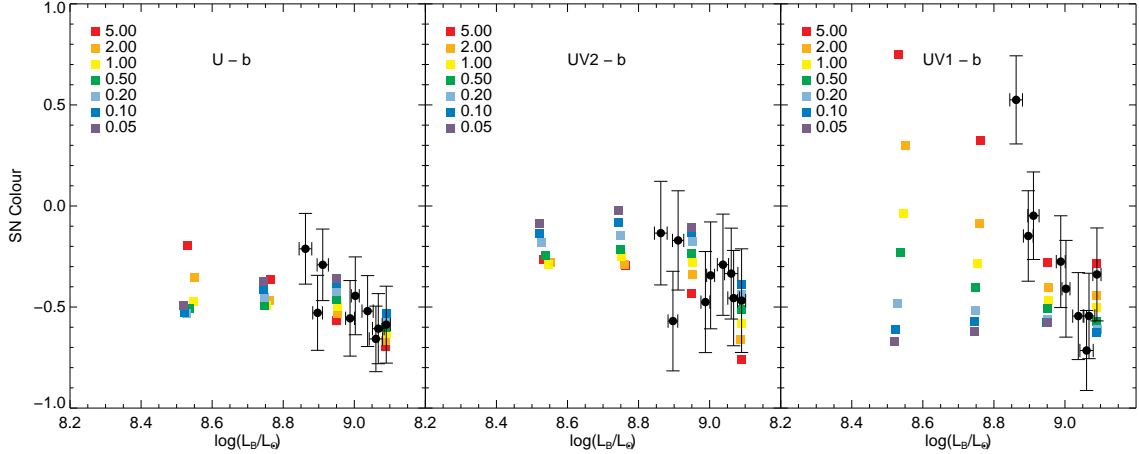}
\caption{UV colours for the models and the data.  The models are shown as the coloured squares and the data in the black filled circles.}
\label{fig:uv_colours}
\end{figure*}

We first examine the broadband UV colours at maximum light as first examined in
\citetalias{Ellis08}.  Their study used Bessel U filter as well as three top-hat
filters they defined in the UV and optical (UV1, UV2 and a normalisation filter
b; see Figure \ref{fig:model_lum}).  We do not perform any correction for
supernova colour and remeasure the U, UV1, UV2, b magnitudes for the models and
the observed spectra.  The results are plotted in Figure \ref{fig:uv_colours}.

This shows that the broad trends observed within the \citetalias{Ellis08} data
at maximum are replicated in the models.  The filters U and UV2 do not show any
strong trend with colour and $B$-band luminosity.  The dispersion between models
of different metallicity is small, and the trend between metallicity and colour
is not always linear.  The UV1 filter shows that for the higher metallicity
models the UV colour can be large for supernovae with lower luminosities.  In
general, the dispersion in the UV1 filter, which is the bluest of the three, is
the largest. Given the smaller dispersion in the models and the large errors in
the data, it is not possible to attribute the change in colour to metallicity,
although this may become possible with more, better data. In the UV1 filter, the
evidence that our less luminous SNe come from regions of higher metal content is
stronger. 

As predicted above, 04D1sk is the reddest object for all three colours.  It
appears particularly red in the UV1-b bands implying that  its ejecta are particularly metal-rich.

\begin{figure*}
\centering
\includegraphics[width = \columnwidth]{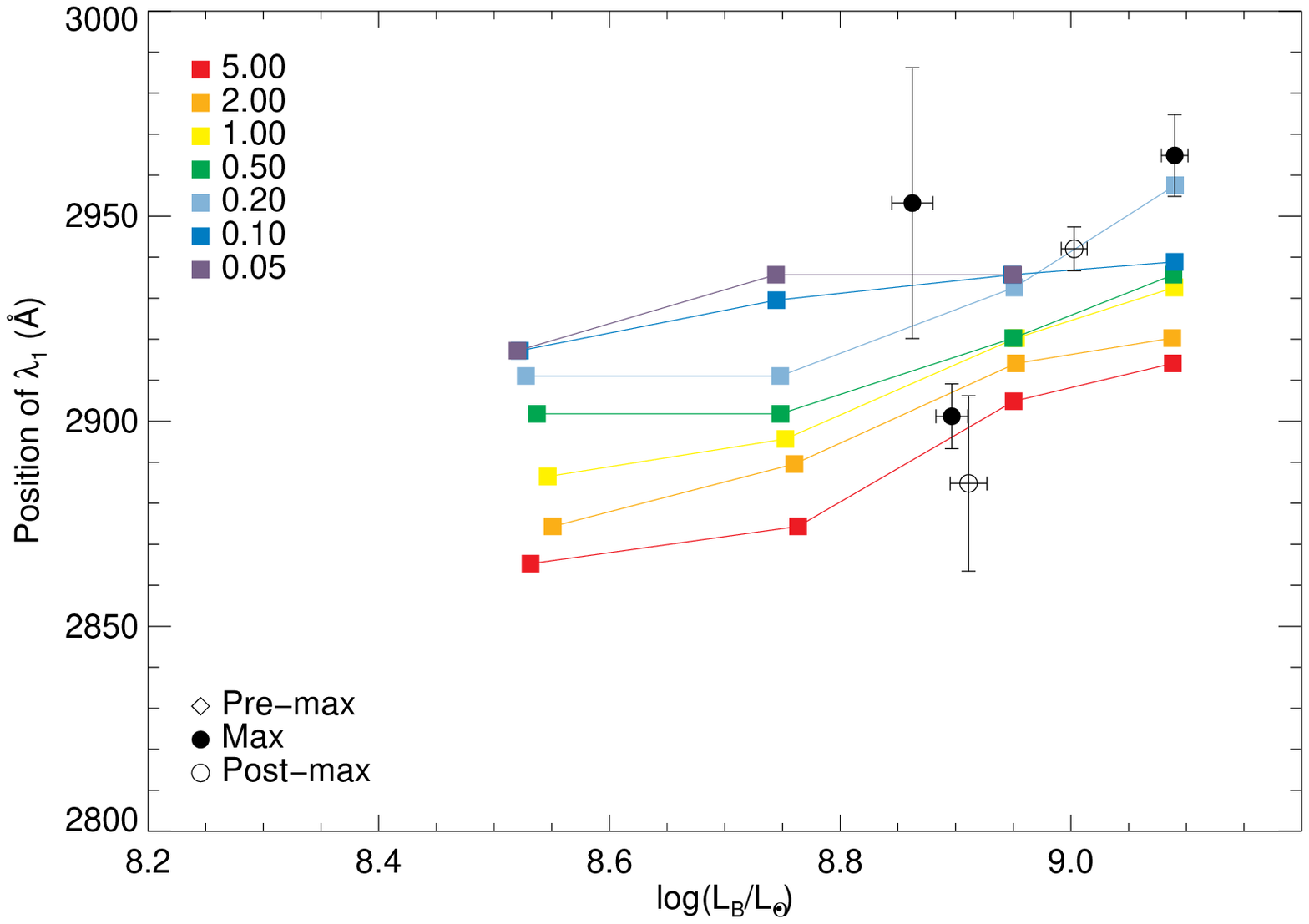}
\includegraphics[width=\columnwidth]{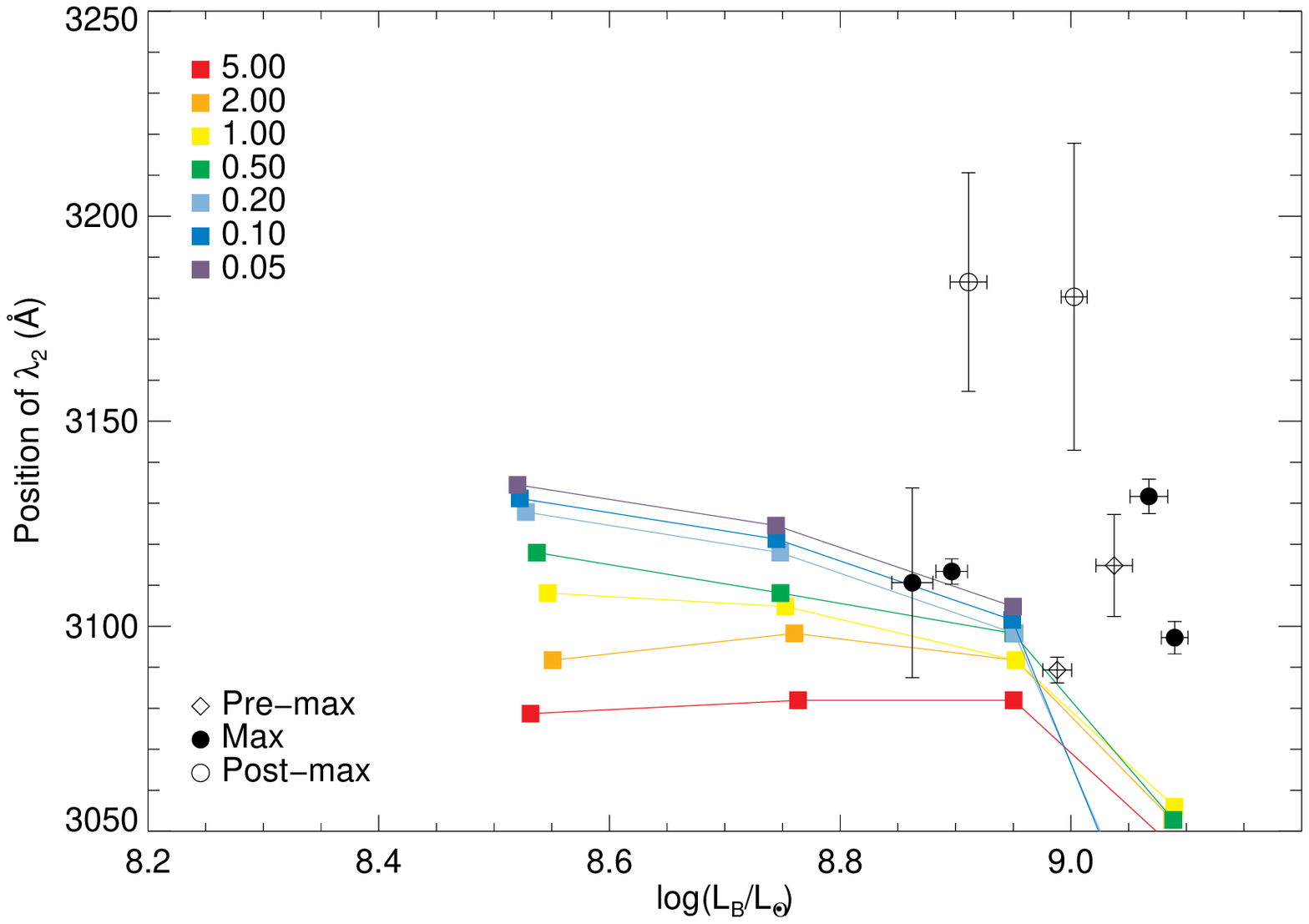}
\caption{The UV emission features $\lambda_1$ (left) and $\lambda_2$ (right) from \citetalias{Ellis08} plotted against $B$-band luminosity for both observed data and values measured from the models.  The different colours denote the different metallicities in the model spectra.  The observed data is divided into sub-samples based on the rest-frame phase of the spectra}
\label{fig:lambda1_lambda2}
\end{figure*}

 \subsection{UV Emission Feature Diagnostics}

Another property examined by \citetalias{Ellis08} is the position of two UV
features described as $\lambda_1$ and $\lambda_2$ positioned at $\approx
2900$\AA\ and $\approx 3100$\AA\ respectively.  The variation of these peaks
with metallicity can clearly be seen in our models in Figure
\ref{fig:model_met}. The reason for this shift can simply be found in the fact
that if the metal content is higher lines are effective at higher velocities,
and hence absorb at bluer wavelengths, progressively shifting the reemission
peaks towards the blue. 

We use the same Gaussian-fitting method as described in \citetalias{Ellis08} to
measure the positions of $\lambda_1$ and $\lambda_2$ in the model data as well
for the observed spectra, although we first apply some smoothing to the observed
data.  As we only have model spectra at one phase (maximum) we choose to further
sub-divide this sample due to the steep observed time-dependence of the position
of these features \citepalias{Ellis08}.  These comparisons are shown in Figure
\ref{fig:lambda1_lambda2} where instead of plotting $\lambda_1$ or $\lambda_2$
against phase, we plot them against $L_B$.  The observed data are divided as
$-3\leq \mathrm{phase} < 1$\,days (pre-max; open diamond); $-1 \leq
\mathrm{phase} < 1$\,days (max; filled circle); $1 \leq \mathrm{phase} \leq
3$\,days (post-max; open circle).

From the left plot in Figure \ref{fig:lambda1_lambda2} we see that at higher
luminosities there is a lower dispersion in the values of $\lambda_1$ in the
models.  The models show that $\lambda_1$ is roughly constant with luminosity in
the lower metallicity models.  We also see an approximately constant value of
$\lambda_1$ within the observed data for all but 2 of the spectra with no strong
phase-dependence.

Figure \ref{fig:lambda1_lambda2} (right plot) shows that the position of the synthetic 
$\lambda_2$ peak emission varies strongly with both luminosity and metallicity
(see above).  The observed data at $\pm1$\,day agrees with the measurement in
the models and the strong phase-dependence of the position of this feature is
observed.  At the highest luminosity, the position of the feature is strongly
blueshifted out of the range of measurements found by \citetalias{Ellis08}. 
This is because these features are strongly affected by the supernova velocities
(see below) and at the highest luminosities the widths of $\lambda_1$ and
$\lambda_2$ are very broad.  This causes the two peaks to merge making the
identification of $\lambda_2$ difficult as the features are not Gaussian in
nature (see Figures \ref{fig:model_lum} and \ref{fig:model_met}).

\subsection{UV Flux Ratio}\label{sec:r_uv}

\begin{figure}
\centering
\includegraphics[width = \columnwidth]{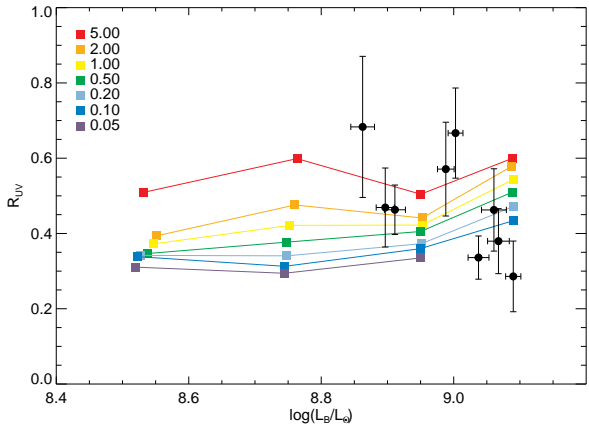}
\caption{The measured $R_{UV}$ for both models and observed data.  The different colours represent the different values of $\eta$ for a given luminosity.}
\label{fig:r_uv}
\end{figure}

We can use our data to measure the UV flux ratio of \citet{Foley:2008fx},
$R_{UV}$, which is defined as

\begin{equation}
R_{UV} = \frac{f_{\lambda}(2770\textrm{\AA})}{f_{\lambda}(2900\textrm{\AA})}\label{eq:ruv}
\end{equation}

\noindent where $f_{\lambda}(2770\textrm{\AA})$ and 
$f_{\lambda}(2900\textrm{\AA})$ are the median fluxes in bands of size
$\pm20$\,\AA\ centred at these wavelengths.  The results are plotted in Figure
\ref{fig:r_uv}.

Figure \ref{fig:r_uv} shows that for a given metallicity ($\eta$), the value of
$R_{UV}$ is approximately constant.  The model trends agree broadly with the
values of $R_{UV}$ measured from the observational data, but in this region of
the spectrum the observed data is strongly affected by noise. We could only
reproduce the correlation between $L_{B}$ and $R_{UV}$ as would be expected from
the results in \citet{Foley:2008fx} if the brighter observed SNe Ia are in more
metal-rich hosts.  Instead the models show that the diversity is driven almost entirely
by the difference in metallicity.   Some trend in seen in the data and the
strength of this correlation is $2.4\sigma$ using a Spearman's Rank correlation
statistic assuming all points are equally weighted.  We do not conclude anything
from this because of the small sample size.

We have also performed an analysis of $R_{UV}$ after performing a
colour-correction to the supernovae using the SALT2 colour law \citep{salt2}. 
The change in the measured value of the ratio with this correction in the order
of 0.005 and so much less than the $1\sigma$ error on the $R_{UV}$ measurement
based on the noise of the data.  We have also examined what would happen should
the host subtraction of the galaxy as described in \citetalias{Ellis08} be
incorrect.  We found that addition/subtraction of any galaxy template used in
\citetalias{Ellis08} with a $u^*$-band flux of up to 20\% that of the supernova
would not change the $R_{UV}$ measurement by more than the $1\sigma$ error.  As
such we do not believe our $R_{UV}$ measurements would be strongly affected by
any inaccuracies in the host subtraction.

\section{Discussion}\label{sec-discussion}

\subsection{UV Colours}\label{sec:disc_colours}

We broadly replicate the wide range of colours observed in the UV with our
models.  We also show that dispersion increases towards the blue.  This is due
to the increasing extent of the metal line-blanketing which affects the whole of
the UV region reflected in the opacity of the SN ejecta increasing by 3 orders
of magnitude from 4000\AA\ to 2000\AA\ \citep{Sauer:2006gj}.  This increase in
dispersion is also consistent with the SWIFT data results \citet{Brown:2010cd}

The UV colours are the best diagnostic we have featured for replicating
properties seen by the whole observed data sample.  This is because they measure
over wide wavelength ranges and so are less affected by variations caused by the
velocities in the models not matching the high-z data precisely.  

\subsection{UV Peaks}\label{sec:disc_l1l2}

The two features designated as $\lambda_1$ and $\lambda_2$ in
\citetalias{Ellis08} are not emission features in the traditional sense. The
sheer number of species absorbing in these regions makes this extremely
unlikely.  The reason for the peaks in flux at $\approx2900$\,\AA\ and
$\approx3100$\,\AA\ could be due to two things, or most likely a combination of
both: a "window" in opacity at these wavelengths which allows more of the
photons emitted from the photosphere to escape, or reverse fluorescence from
species in the outer layers of the ejecta are generating these blue photons
\citep{Mazzali:2000va}.  We can use the output of the model code to investigate
what happens to photon packets emitted from the photosphere to try and
differentiate between the two.

Firstly, we look at what happens to packets emitted from the photosphere in a
region of $\pm40$\,\AA\ around the wavelengths of the  measured values of
$\lambda_1$ and $\lambda_2$.  We find that at these wavelengths, for both
features and at all luminosities and metallicities, between 50\% -- 80\% of the
packets are reabsorbed back into the photosphere through backscattering.  Of the
small number of packets left, they are mostly re-emitted at redder wavelengths. 
This means that an opacity window at low velocities can be excluded as the
dominant effect.

\begin{figure*}
\centering
\includegraphics[width = \textwidth]{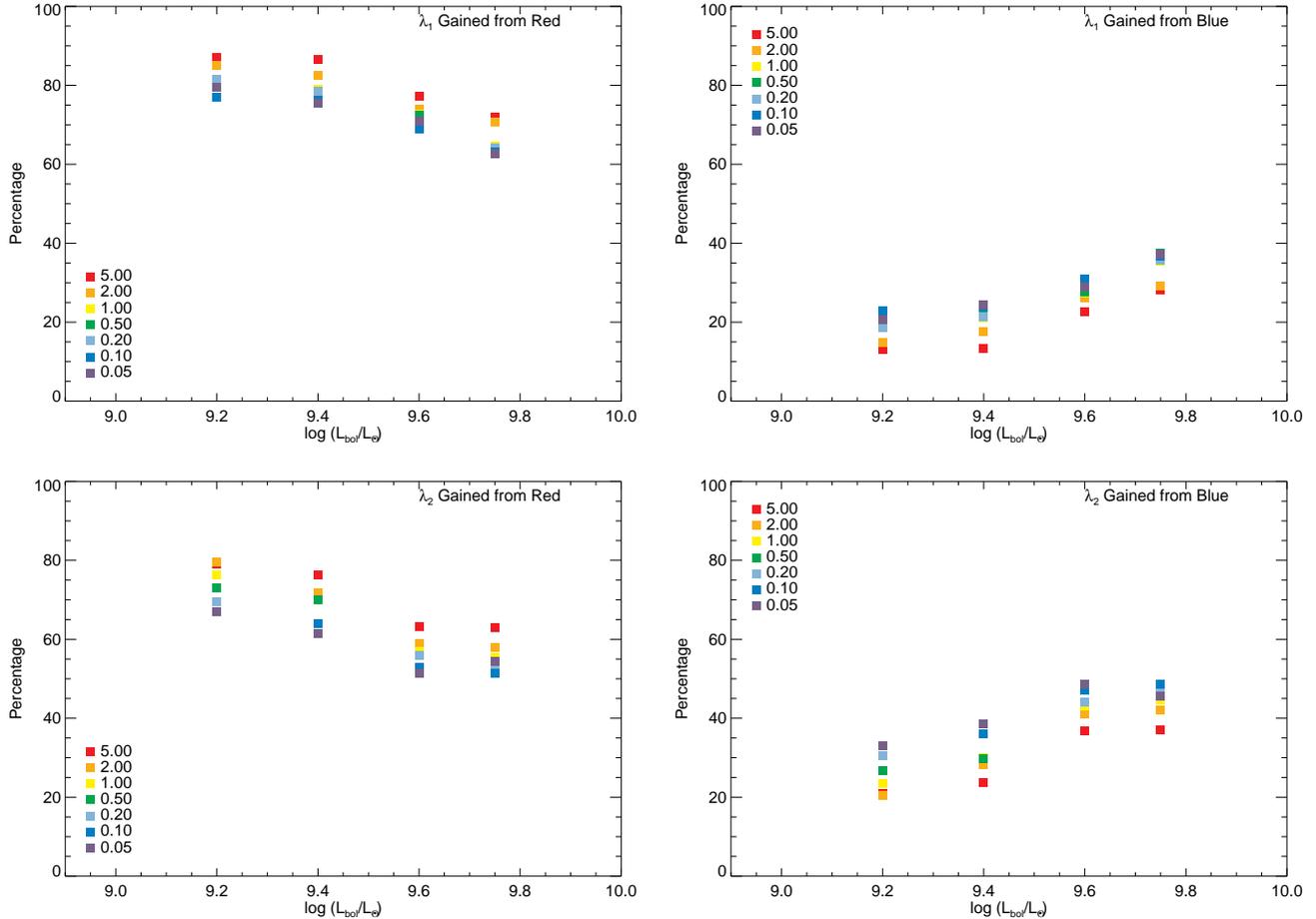}
\caption{The percentage of packets gained in the $\lambda_1$ (top) and $\lambda_2$ (bottom) regions which were originally emitted at redder (left) or bluer (right) wavelengths.}
\label{fig:packets_won}
\end{figure*}

We can also look at photon packets which were originally emitted in other
regions of the spectrum, absorbed in some line transition and finally re-emitted
within the $\lambda_1$ and $\lambda_2$ regions.  This is illustrated in Figure
\ref{fig:packets_won}.  We see here that for $\lambda_1$ by far the most packets
are gained from the red in reverse fluorescence processes, although this
percentage decreases as luminosity increases.  The situation is similar for the
$\lambda_2$ feature, although by the time we reach our highest luminosity model,
the percentage is roughly 50\% from the blue and 50\% from the red. Reverse
fluorescence was already mentioned as being primarily responsible for the
emerging UV flux by \citet{Mazzali:2000va}: photons are fluoresced back to the
UV at high velocities, and thus they can escape in low-oacity environments.

If we also look at this measurement in terms of the \emph{number} of packets gained
rather than the percentage of packets, we actually see that for both features the
number gained from the blue increases with luminosity, which is to be expected
because of the increase in temperature.  The number of packets gained via
reverse fluorescence, meanwhile, does not increase in the same way.  For
$\lambda_2$  the number increases in the lower two models and after that remains
constant meaning that the lines in the optical have become saturated and this is
limiting the number of photons re-emitted in this region.  The number of packets
for $\lambda_1$ does not saturate and so this is still affected by the
abundances of the IME and IGE which absorb photons in the optical. 

We can also examine which ions cause the processes that shift flux into the $\lambda_1$ and $\lambda_2$ regions.  We
find that for both features, the predominant species which
reverse fluoresce are IME, particularly \ion{Mg}{ii}, \ion{Si}{ii} and
\ion{S}{ii}, with some contributions from the Fe-group elements, notably
\ion{Cr}{ii}.  As these features are dependent on abundances of elements they
will be affected by any differences between the velocity structure and the
abundance stratification between the models and the data.  This can account for
some of the discrepancy we see between the positions of the features.

\subsection{UV Ratio}\label{sec:disc_ruv}

It has been established that a source of UV flux is a process of reverse
fluorescence where red photons are absorbed and re-emitted at shorter
wavelengths at high velocties \citep{Mazzali:2000va}.  This means that there
have to be some metals in the material above the photosphere or there would be
no UV flux at all.  As the metal content is increased, the UV flux increases due
to an increasing amount of reverse fluorescence; however this is in competition
with an increased absorption of the UV photons by the metals themselves.   The
way this occurs in the regions of $f(2770)$ and $f(2900)$ determines how
$R_{UV}$ changes with metallicity at constant luminosity.   

\begin{figure*}
  \centering
  \includegraphics[width=\textwidth]{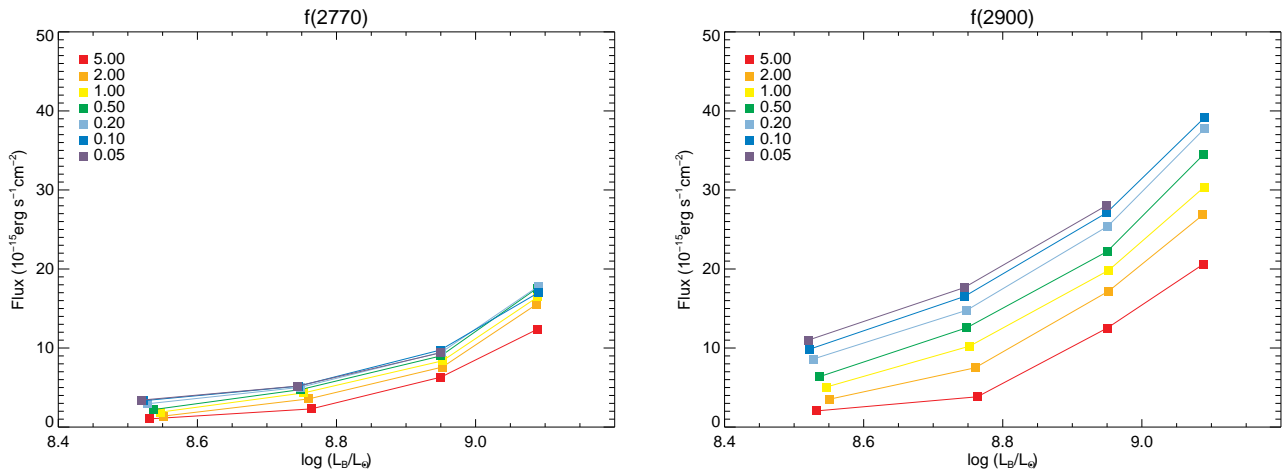}
  \caption{Median flux values in the two regions probed by $R_{UV}$ for our model $(L,\eta)$ sequences.}
  \label{fig:median-flux}
\end{figure*}

We have plotted the values $f(2770)$ and $f(2900)$ in our models in Figure
\ref{fig:median-flux}.  In all cases, we see that the higher $\eta$ models have
less flux as there are more metals absorbing in these bands.  In both bands we
see a dispersion with metal content, which is larger for $f(2900)$.  This shows
that the variation in $R_{UV}$ is driven more by the variation in the $f(2900)$
band.

We used the MC code to identify the dominant absorption lines in the regions
probed by $R_{UV}$ looking at ions with a large absolute change in absorption
strength between $\eta = 0.05$ and $\eta = 5$ (Table \ref{tab:strong_species})
or ions which have large mean absorption strengths averaged over the metallicity sequence
(Table \ref{tab:strong_species2}).  We can do this analysis for $R_{UV}$ because
of the small wavelength ranges for $f(2770)$ and $f(2900)$.  No single element
appears to be responsible for the change, although the $f(2900)$ region appears
to be dominated by variations in the \ion{Cr}{ii} lines.  We see that in the higher luminosity
models, the dominant species change from singly-ionised to doubly-ionised
reflecting the increase in temperature in the atmosphere.  Figure
\ref{fig:median-flux} explicitly shows that effect of varying $\eta$ has on the
two spectral bands.

\begin{table}
  \centering
  \caption{The species which show the largest absolute change in line 
  strengths with metallicity in the two regions probed by $R_{UV}$.  }
  \label{tab:strong_species}
   \begin{tabular}{c l l}
  \hline
  $\log(L/L_{\odot})$ & $f(2770)$ Region & $f(2900)$ Region \\
  \hline
    9.2 & \ion{Fe}{ii} & \ion{Cr}{ii}\\
  9.4 & \ion{Fe}{ii} & \ion{Cr}{ii}\\
  9.6 &  Singly-ionised metals & \ion{Cr}{ii} \\
  9.75 & Doubly-ionised metals &  \ion{Cr}{ii}\\
  \hline   
   \end{tabular}
 \end{table}

\begin{table}
  \centering
  \caption{The species which show the largest mean line strengths across all 
  metallicities in the two regions probed by $R_{UV}$.  ``--" marks mean that the region contains a mix of singly- and doubly-ionised 
  species of a variety of elements.}
  \label{tab:strong_species2}
   \begin{tabular}{c l l}
  \hline
  $\log(L/L_{\odot})$ & $f(2770)$ Region & $f(2900)$ Region \\
  \hline
  9.2 & \ion{Fe}{ii} & Singly-ionised metals\\
  9.4 & \ion{Fe}{ii} & Singly-ionised metals\\
  9.6 & Singly-ionised metals & -- \\
  9.75 &  Doubly-ionised metals &  Doubly-ionised metals\\
    \hline   
   \end{tabular}
 \end{table}

\subsection{Focus on Individual Elements}\label{sec:elements}

As well as creating sequences of models where the multiplier $\eta$ affects all
elements with $Z>20$, we can use $\eta$ to vary one element while keeping all
others fixed.  This allows us to examine whether this one element is the cause
of any of the observed effects we see.  Should any of this be caused by one
element alone this would give us an important diagnostic to examine the physics
of the supernova explosion itself.  The species for which this analysis was
carried out are chromium, stable iron, manganese, \nickel\ and titanium which
are the elements with species with the strongest features in the near-UV.  We
summarise the effect of individual elements below:

\begin{itemize}
\item \textbf{UV Colours:}  Ti and Mn have no effect on the UV colours.  Stable iron has little effect, except in the UV1-b colour where it shows strong non-linear trends with iron content at a given luminosity.  The chromium and \nickel\ model sequences both show a large dispersion in the UV1-b colour as seen for all metals in Figure \ref{fig:uv_colours} above.\\
\item \textbf{UV Features:}  The position of $\lambda_1$ is strongly dependent on the increased amount of Cr in the spectrum.  This is expected from looking at which elements have caused the reverse fluorescence in this feature.  The position of $\lambda_2$ is affected by the IGE produced in nucleosynthesis i.~e.~ iron, nickel and chromium, but not manganese.  We see that these IGE are small contributors to the reverse fluorescence flux which is dominated by IME, so $\lambda_1$ and $\lambda_2$ will be very sensitive to small changes in the relative abundances of all these elements. \\
\item \textbf{UV Ratio:}  The sequences with varying stable iron and chromium content from nucleosynthesis  all show large variations at constant luminosity.  This is expected as their ions were identified in Tables \ref{tab:strong_species} and \ref{tab:strong_species2}.  There is diversity shown for all of the element sequences although to a lesser degree.  This implies that this feature could vary significantly between supernovae and be dependent on velocity (which absorptions are shifted into the bands); temperature or luminosity affecting ionisation balanaces; and the abundances of the individual elements relative to each other which may not be the same as for SN\,2005cf.
\end{itemize}

\noindent The plots detailing these results can be found in the Appendix.

\begin{figure*}
  \centering
  \includegraphics[width=\textwidth]{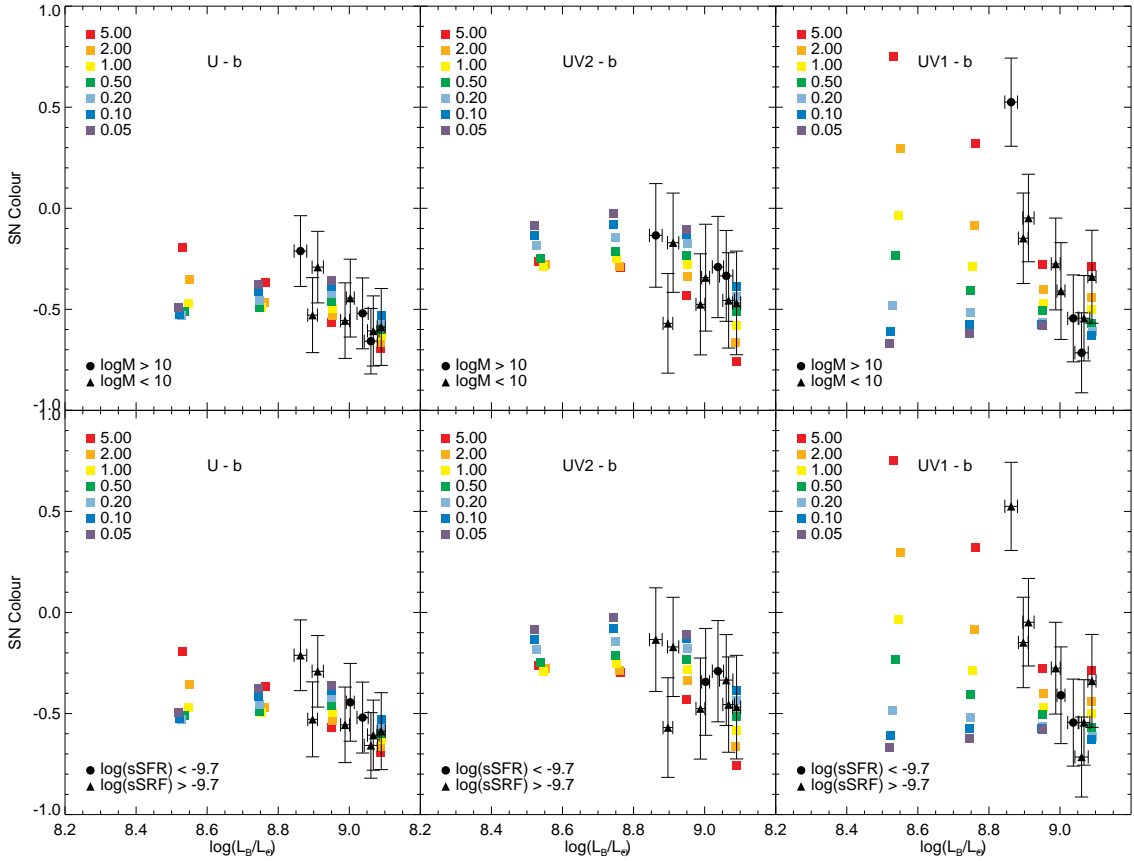}
  \caption{UV colours in the models and the observed spectra with the observed objects from \protect\citetalias{Ellis08} divided into two sub-samples based on their host mass (top) and the host specific star formation rate (bottom). The values of the specific star formation rate (sSFR) are from \citet{Sullivan:2010ez}. The model sequences for different values of the metal content in the SN outer layers ($\eta$) are again plotted in different colours.}
  \label{fig:uv_colours_hostprops}
\end{figure*}

 \subsection{Links to Other Observables}

\subsubsection{Host Metallicity}\label{sec:hostprops}

It is not possible to use our models to explain directly the observed trends with host properties.  This is because our method of generating spectra does not separate out the effect of an increased metal content in the outer layers of the SN ejecta due to progenitor or environmental metallicity from any potential upmixing of elements which are burnt during the SN explosion.

Hints on whether the environmental metallicity plays a major role for the UV
colours may directly be obtained from the observations by subdividing the
observed sample according to metallicity indicators. In Figure
\ref{fig:uv_colours_hostprops}, we have done this, with the host galaxy mass and
its specific star formation rate, sSFR, as measured in \citet{Sullivan:2010ez}
as indicators.  More massive galaxies with lower star-formation rates should
generally be old, red ellipticals which have a high metallicity. In the upper
panel of Figure \ref{fig:uv_colours_hostprops}, there are two SNe in massive
(high-metallicity) galaxies which are found to have low-$R_{UV}$ /
high-luminosity; however, there is also one SN  in a high-mass host which shows
an inverse tendency so we do not see any trend with host mass.  Using the
sSFR as a criterion (Figure \ref{fig:uv_colours_hostprops}, lower panel), we see
a similar result.
Fr

\subsubsection{Supernova Light-curve Standardisation}\label{sec:standarise}
\begin{figure}
  \centering
  \includegraphics[width=\columnwidth]{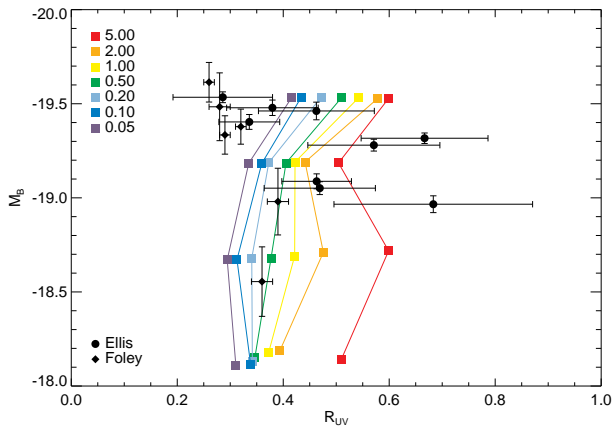}
    \caption{$B$-band magnitudes plotted against $R_{UV}$ for the models in this paper and objects investigated in \protect\citet{Foley:2008fx} and \protect\citetalias{Ellis08}. No correction for light-curve shape or colour has been made to the data.}
  \label{fig:bmag}
\end{figure}

\citet{Foley:2008fx} proposed that an observed correlation between $R_{UV}$ and
the $V$-band magnitude could be used for standardisation of supernova
magnitudes.  In Figure \ref{fig:bmag}, we extend this into the $B$-band and
compare the supernovae in that sample to those from \citetalias{Ellis08} and our
models, where the values of $M_B$ from the models are generated via synthetic
photometry.  For the \citet{Foley:2008fx} data, the $B$-band magnitudes are
taken from \citet{Altavilla:2004hk}.  The low-z data includes the measurement of
$R_{UV}$ nearest to maximum for SN\,1980N, SN\,1981B, SN\,1990N, SN\,1991T and
SN\,1992A.  Figure \ref{fig:bmag} shows that within the luminosity region
populated by SNe Ia which obey the Phillips relation the \citetalias{Ellis08} data do not confirm
the linear correlation suggested in \citet{Foley:2008fx}.  The range and scatter of $R_{UV}$ measurements of the \citetalias{Ellis08} sample appear to have more in common with the supernovae in the more recent \citet{Foley:2012ej} study which uses $M_V$ so cannot be directly compared to these values.

A larger sample of well-observed SNe may be needed to clarify this disagreement.
In any case, $R_{UV}$ is a flux ratio, not a broad-band colour measurement, so
it is very sensitive to slight changes in the UV spectrum, which can be caused
not only by different abundances in the ejecta, but also by small shifts in
velocity.  At 2770\AA, a change of 20\AA\ is equivalent to a change in velocity
of $\approx 2000$\,km\,s$^{-1}$.  This is velocity change which could quite
easily occur in normal SNe Ia.  The measurements will also be strongly influence
by the phase of the supernova as we can expect velocity shifts as observed in
the optical.  The phase-dependence of the $\lambda_2$ is shown in Figure
\ref{fig:lambda1_lambda2}.

The best opportunity to use the UV for light curve calibration is the  strong
trend between $L_B$ and the (UV1 - b) colour as shown in Figure
\ref{fig:uv_colours}.  The b filter corresponds to the optical region where
supernovae are less susceptible to metallicity differences, and the features
maintain a constant morphology despite differences in flux.  Figure \ref{fig:uv_colours} shows that
for the same luminosity the UV changes dramatically with metallicity, while $L_B$ is
basically constant.

The UV1 region changes dramatically with both luminosity and metallicity which
drives the trend seen Figure \ref{fig:uv_colours}.  It is possible that with
optimisation, this colour could be used for lightcurve standardisation.  (U-b)
also shows a linear trend in the observed data as the U filter extends down as
far as 3000\AA, but at lower luminosity the models show the trend is not
monotonic with metallicity.  This would make any standardisation difficult.

\subsection{Future Extensions of This Study}\label{sec:future}

The observational data used in this paper are subject to large errors in both
luminosity and $R_{UV}$.  New observations with higher S/N in the UV would
reduce the errors on $R_{UV}$.  Observations of supernovae with lower luminosities would also enable a thorough comparison with the model data in these ranges.  

In terms of the modelling, we have used only one well-studied supernova as the
basis of our models.  While our spectra appear normal at a variety of
luminosities and metallicities, they are artificial in having been produced by
arbitrary changes to the input parameters.  With more well-studied supernovae
with UV data, we would have more starting points for the creation of more
accurate models covering a wider range of initial luminosities.  This may then
help to produce more accurate replicas of observed supernovae and reduce some of
the disagreement seen between the models and the data.  Velocity changes
particularly affect the single feature and narrow-band measurements so sets of
models which explore different parts of this parameter space would be especially
helpful here.

In the future it may also be possible to do a similar analysis for pre-maximum
spectra where spectral signatures of the unburnt material from the progenitor
are more easily identified in the optical.

 \section{Conclusions}\label{sec:conclusions}
  
We have used maximum-light spectra for a sample of 9 SNe Ia  obtained by
\citetalias{Ellis08} and compared them to a series of 1D models produced by a MC
radiative transfer code in order to test whether the variations observed the UV
region of the spectrum can be explained by metallicity and luminosity changes
alone.  Our model spectra were initially based on one-well observed example
SN\,2005cf.  We summarise our conclusions in the following points

\begin{itemize}
\item Our models replicate well the range of broad-band UV colours seen in the 
observed spectra, showing that the dispersion increases as one progresses to shorted wavelengths.  The (UV1-b) colour appears to depend strongly on metallicity,
and it might have the potential for use in light-curve standardisation because 
it is less sensitive to very small changes in absorption features than 
narrow-band indicators(Sections 
\ref{sec:disc_colours} and \ref{sec:standarise}); however this requires further observational data for testing at lower luminosities.
\item We observe that $\lambda_1$ and $\lambda_2$ can be interpreted as peaks 
caused by the reverse fluorescence of photons into the UV region of the 
spectrum, mainly from intermediate mass elements and chromium 
(Section \ref{sec:disc_l1l2}). Both $\lambda_1$ and $\lambda_2$ move towards the
blue with increasing metal content in the upper layers of the ejecta, but the
effect is highly non-linear and may not be the best metallicity indicator. 
\item We see that $R_{UV}$ in our models has a low dependence on luminosity and 
the effect of the metal content is complex (Section \ref{sec:disc_ruv}). High-z
data do not confirm the relation of \cite{Foley:2008fx}, and so we do not support using this
index for light curve standardization. 
\item We have used high-z Type Ia supernovae spectra from \citepalias{Ellis08} and 
measured observed values of $R_{UV}$ for 9 objects around maximum light.  
We see that the measured values agree with those in the models, although they 
are affected by large errors from noise in the spectra (Figure \ref{fig:r_uv}).
\item We have performed an extensive search using the model spectra in order 
to investigate the possibility that these results are driven by one or two 
elemental transitions, but the results have shown that the UV spectrum is far 
too complicated for this.  In light of this we suggest that the use of 
$R_{UV}$ is not appropriate for light-curve standardisation as the trends with 
metallicity and absolute magnitude are not linear (Section \ref{sec:elements}).
\end{itemize}

 \section*{Acknowledgements}

We acknowledge support from the Italian Space Agency (ASI) under contract 
ASI/INAF n. I/009/10/0 and I/016/07/0.  ESW and SH are grateful for the hospitality of MPA, Garching and the Weizmann Institute of Science, Rehovot during stages of this work.  Joint work by AG and MS is supported by the Weizmann-UK ``making connections'' programme.  Joint work by AG and PAM is supported by a Weizmann Minerva grant.  AG further acknowledges support from the ISF, an ARCHES award, and the Lord Sieff of Brimpton fund.

\bibliographystyle{mn2e}
\bibliography{bib_070512.bib}
\appendix

\section{Single Element Sequences}

\begin{figure*}
  \centering
  \includegraphics[width = \textwidth]{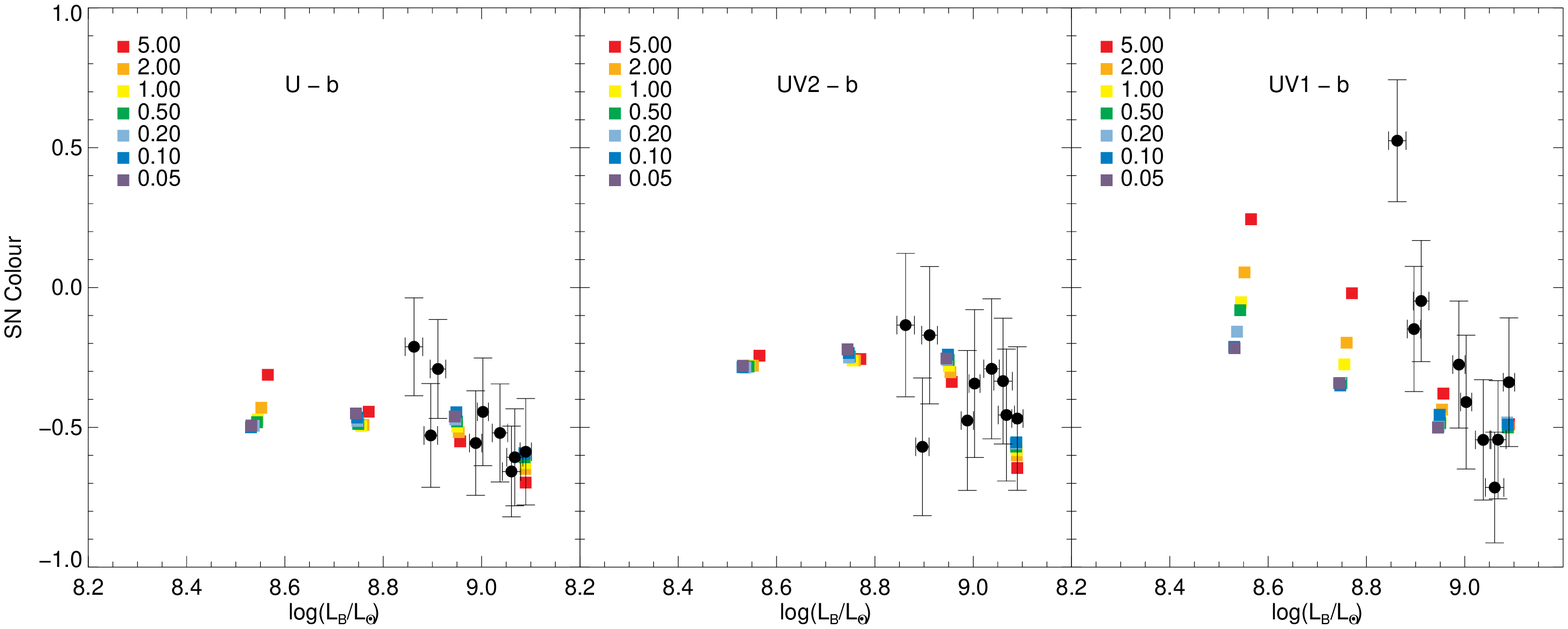}
  \includegraphics[width = \columnwidth]{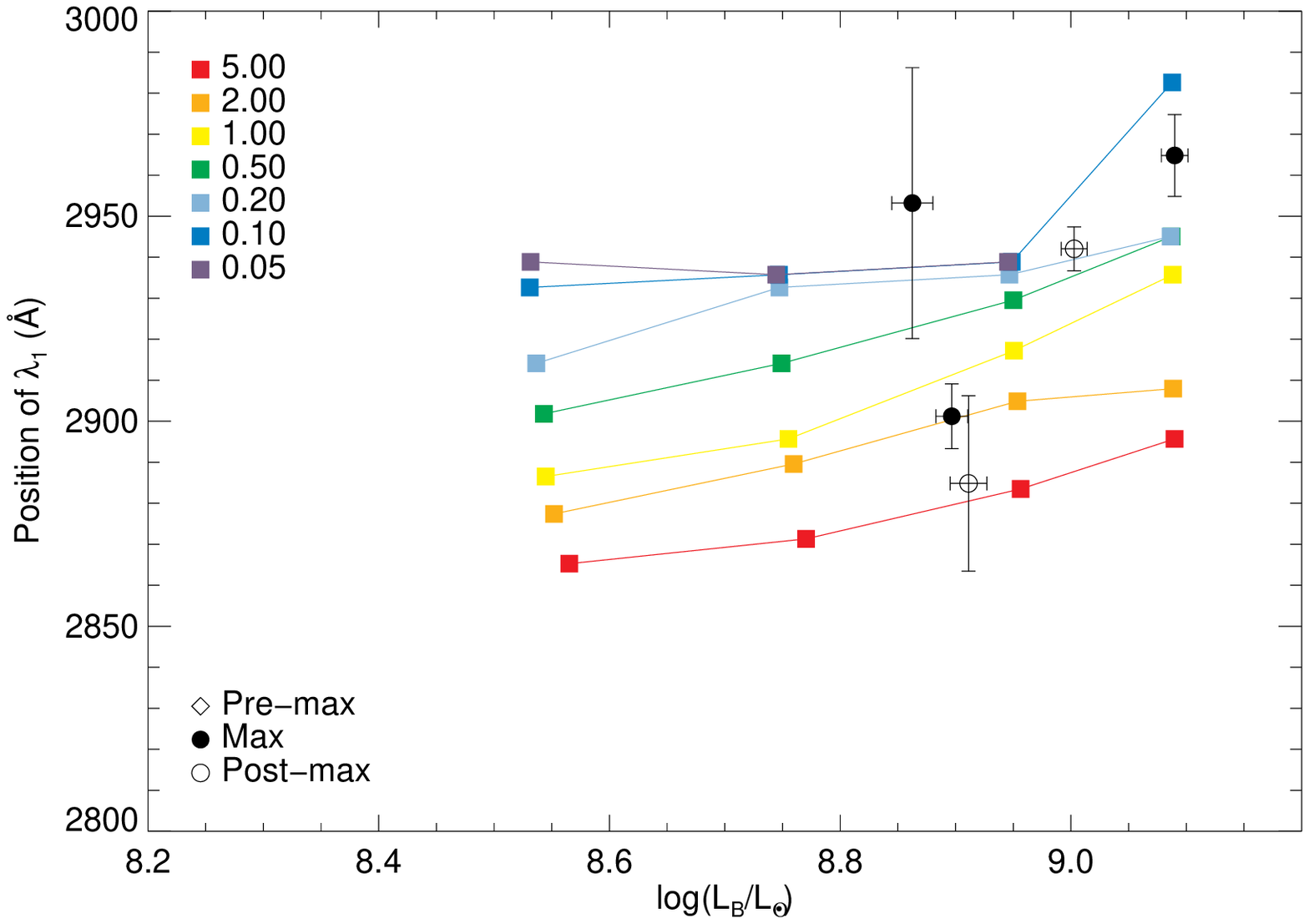}
    \includegraphics[width = \columnwidth]{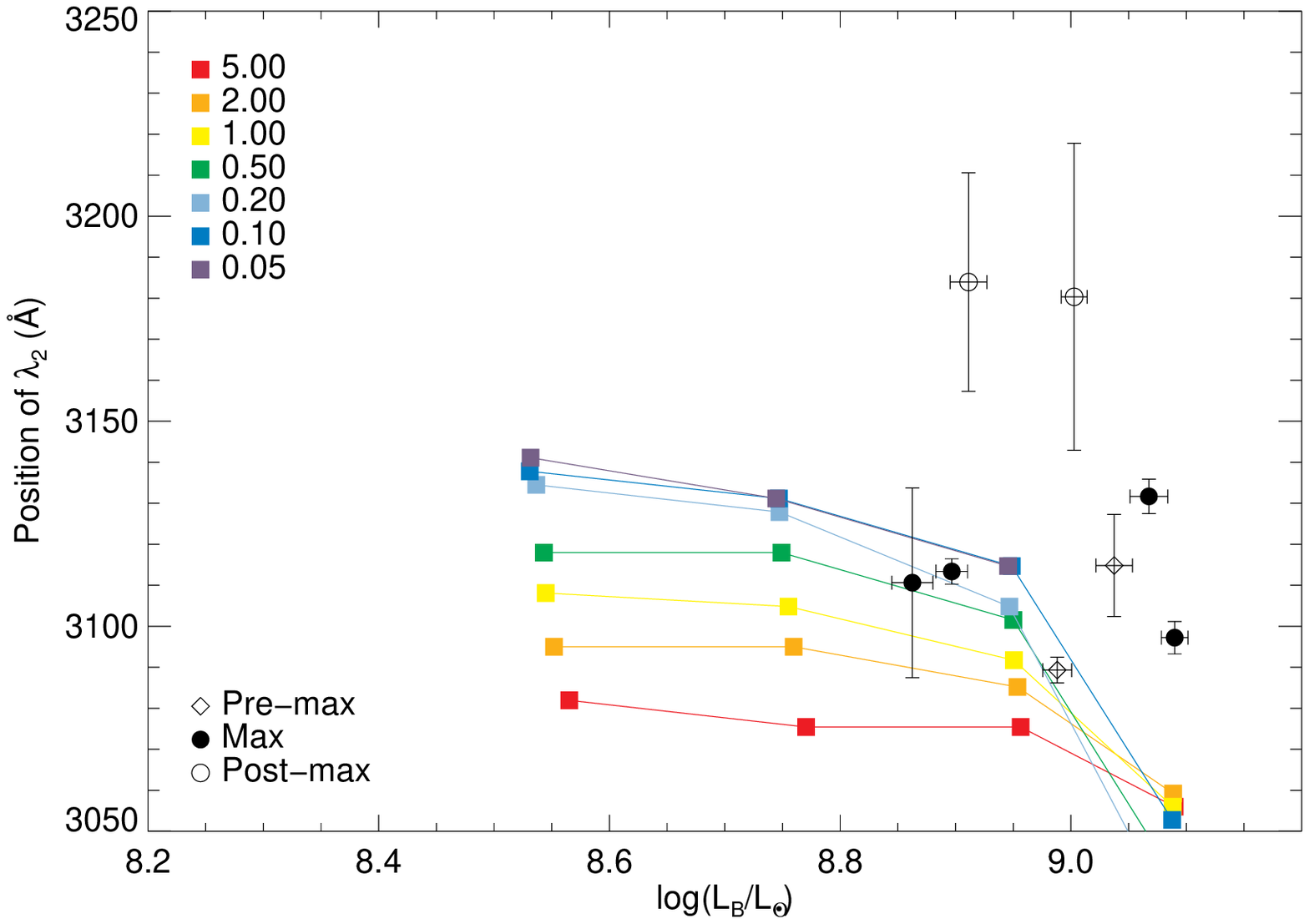}
 \includegraphics[width = \columnwidth]{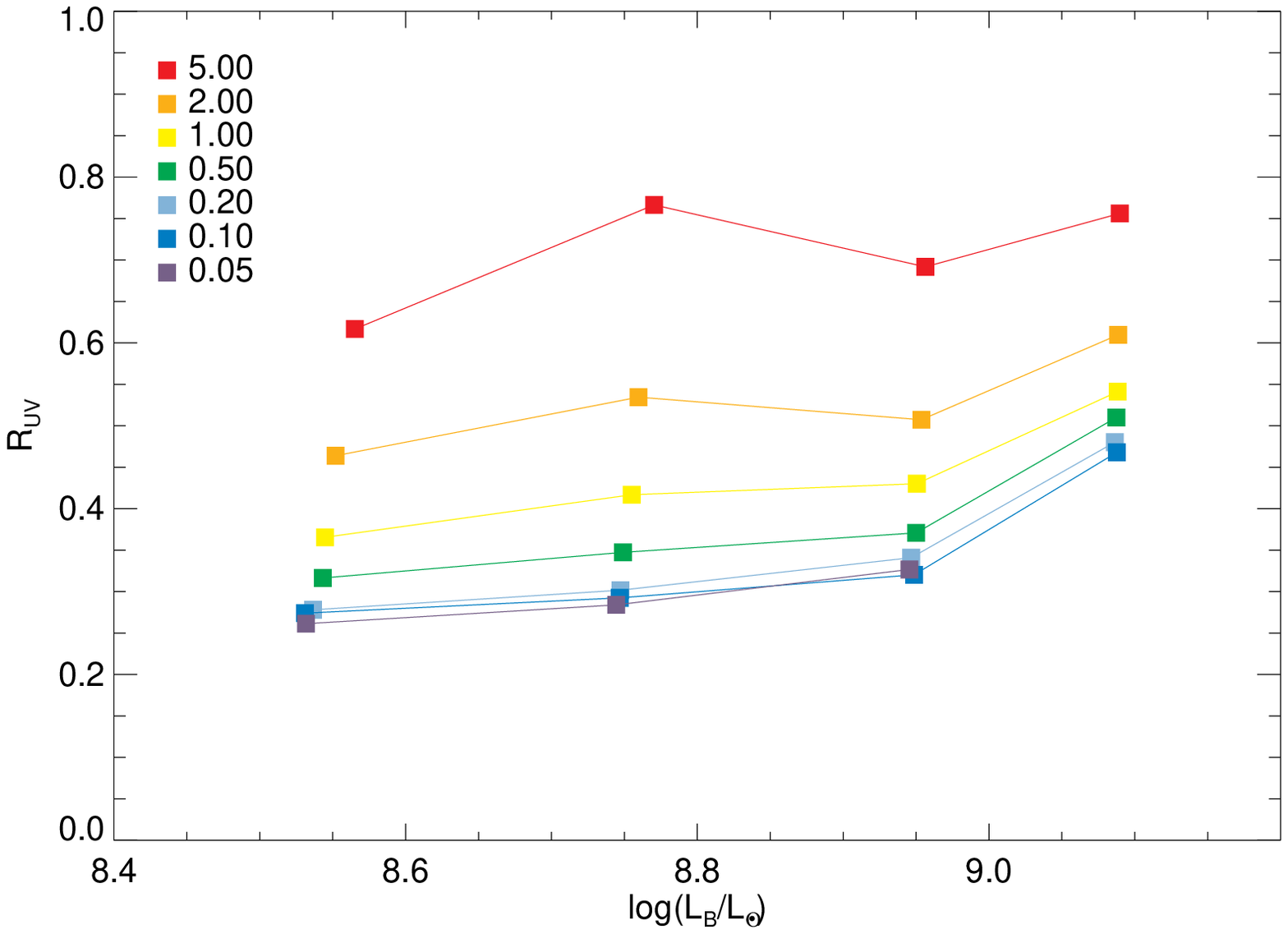}
  \caption{The effect of changing just the amount of chromium in the outer ejecta.  Upper -- UV colours; Middle -- The positions of $\lambda_1$ and $\lambda_2$;    Lower: $R_{UV}$ }
  \label{fig:cr_seq}
\end{figure*}

\begin{figure*}
  \centering
  \includegraphics[width = \textwidth]{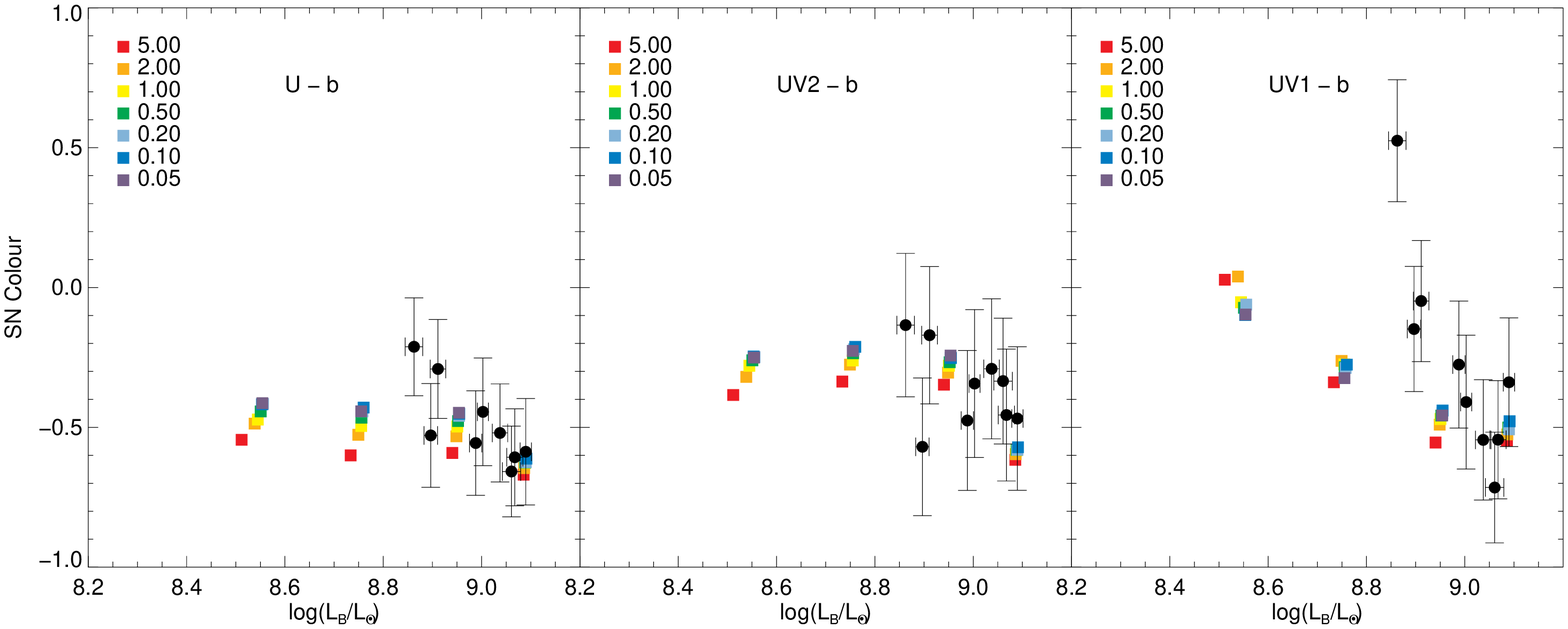}
  \includegraphics[width = \columnwidth]{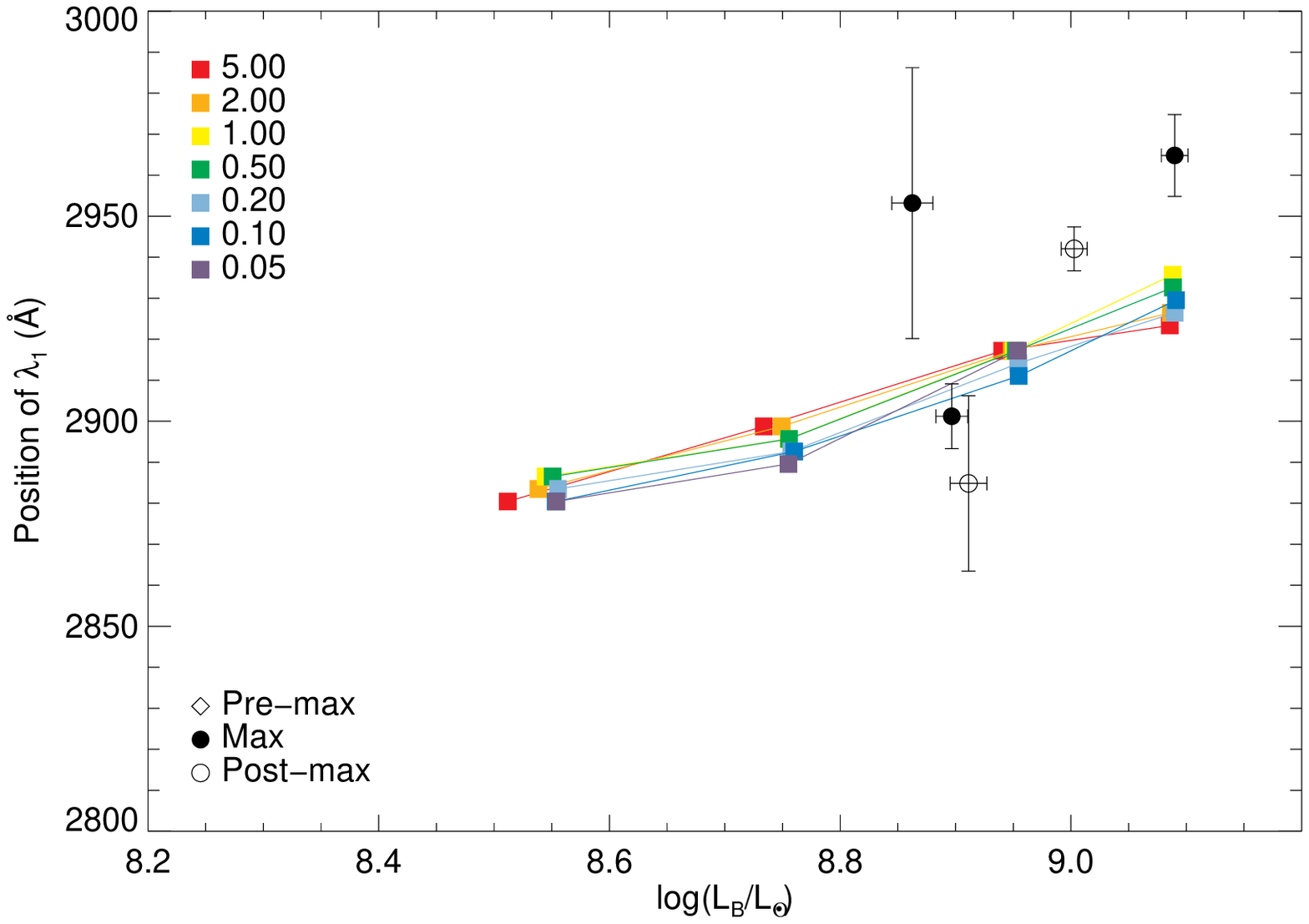}
    \includegraphics[width = \columnwidth]{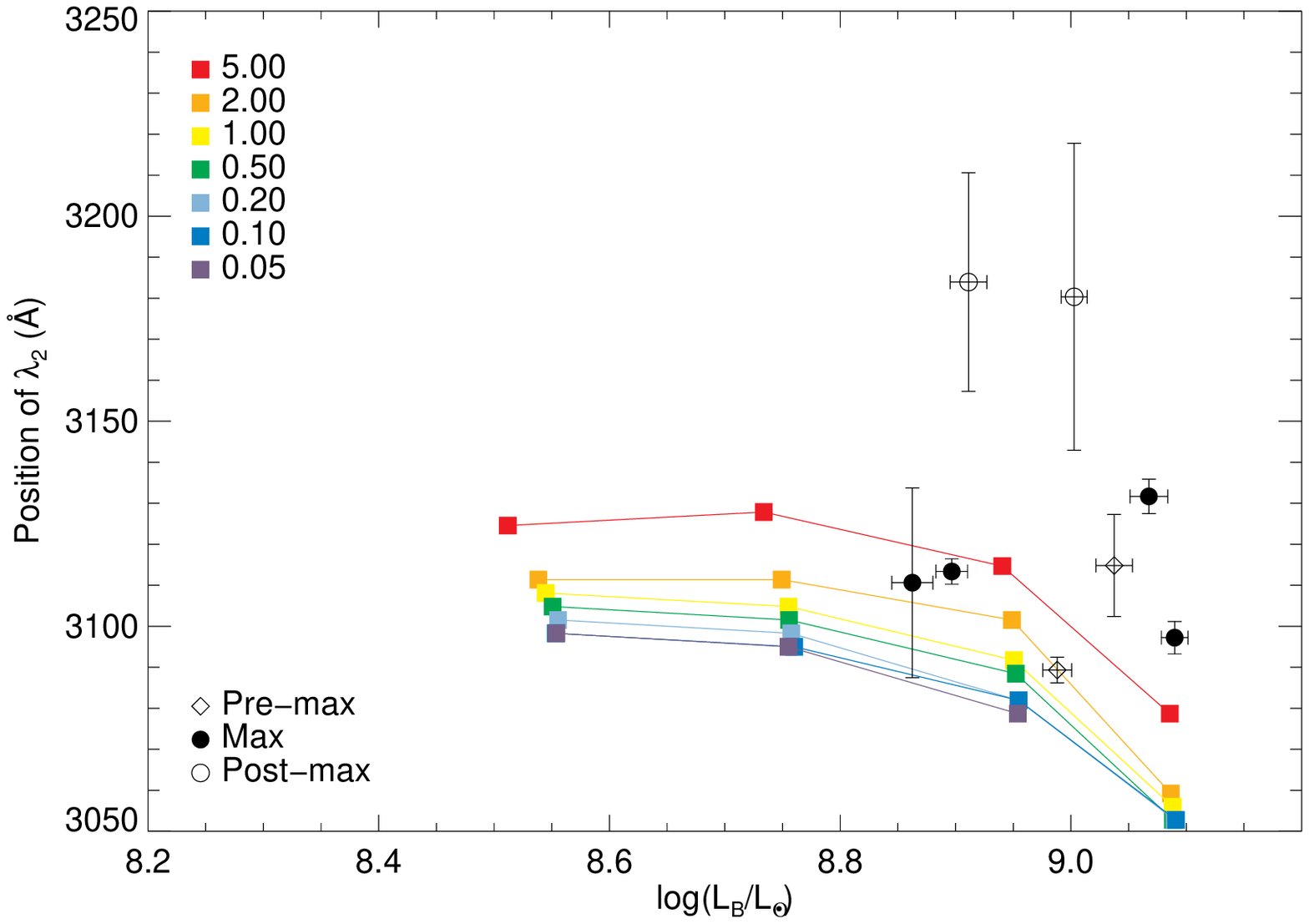}
 \includegraphics[width = \columnwidth]{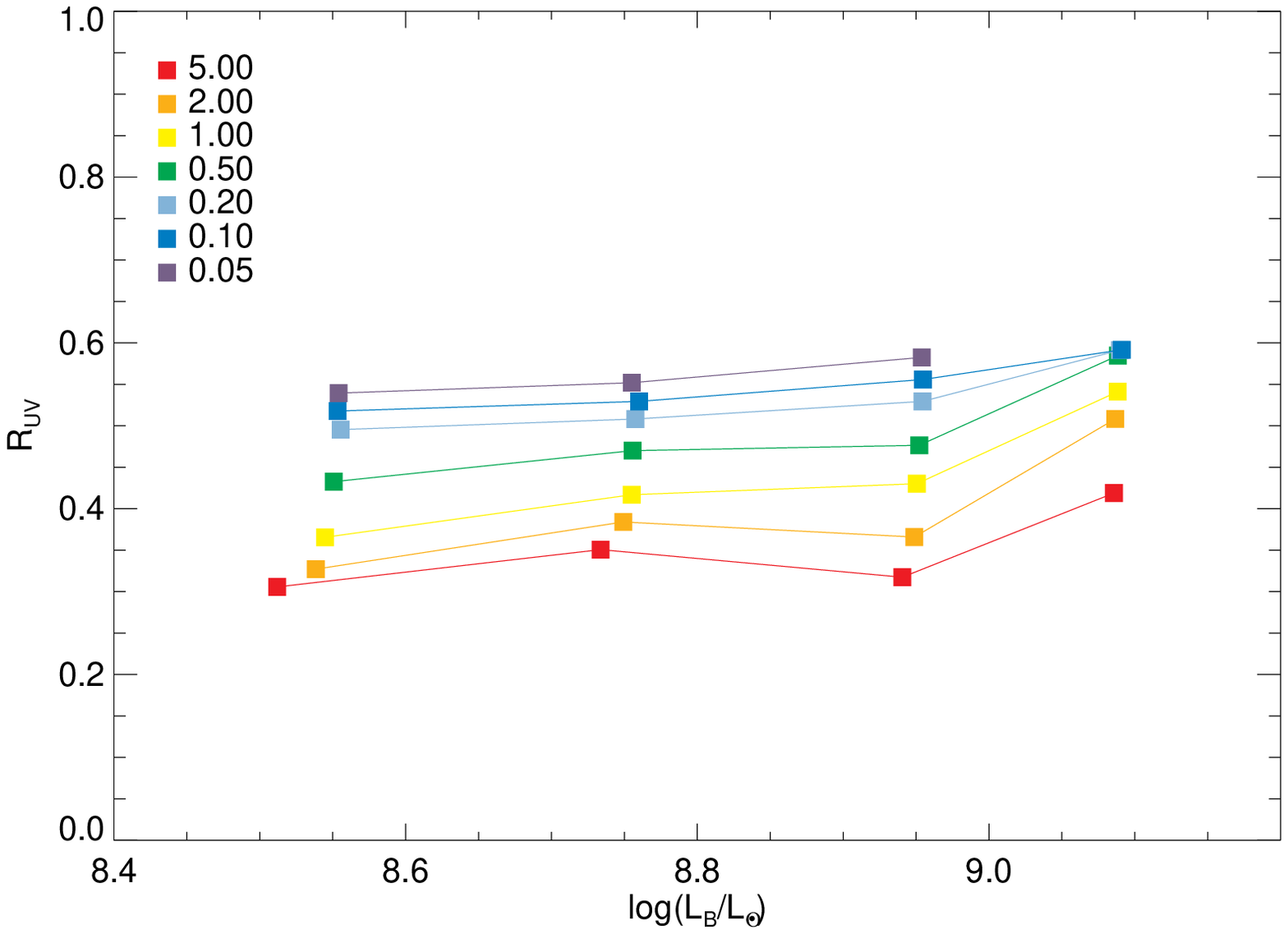}
  \caption{The effect of changing just the amount of stable iron in the outer ejecta.  Upper -- UV colours; Middle -- The positions of $\lambda_1$ and $\lambda_2$;    Lower: $R_{UV}$ }
  \label{fig:fe_seq}
\end{figure*}

\begin{figure*}
  \centering
  \includegraphics[width = \textwidth]{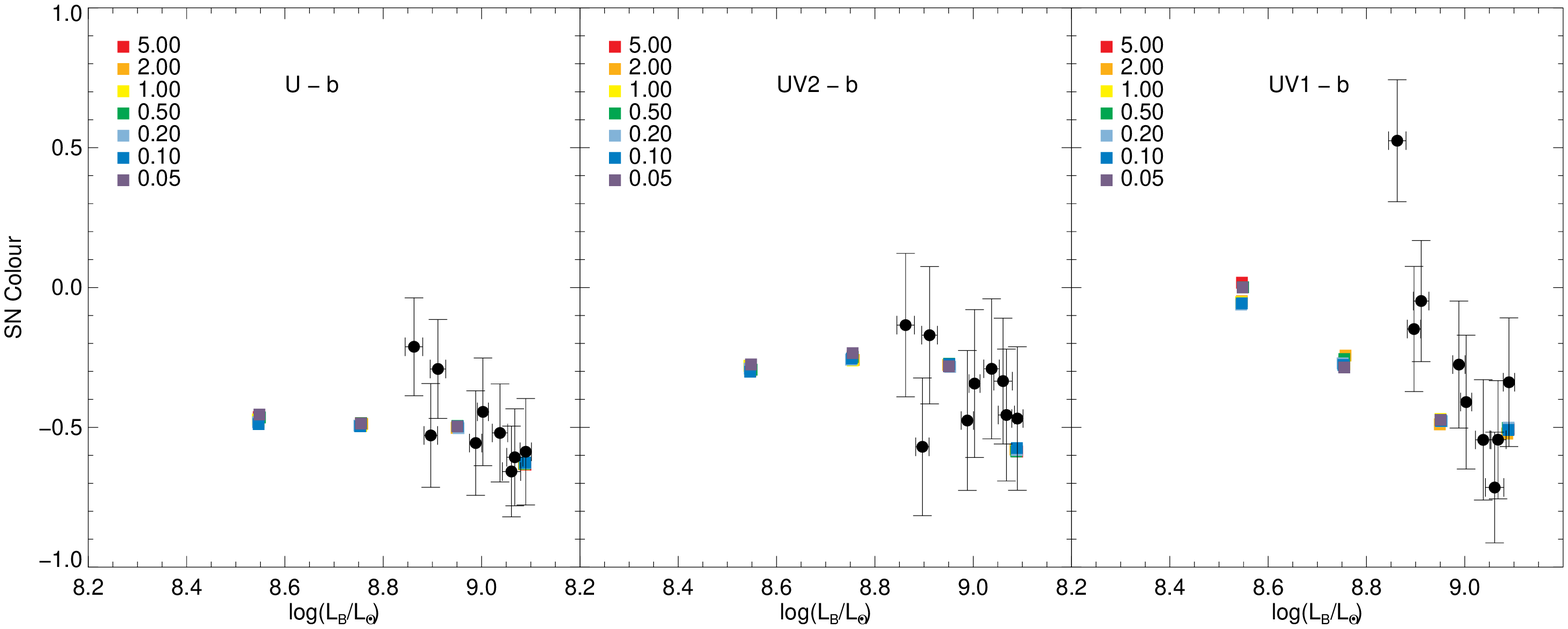}
  \includegraphics[width = \columnwidth]{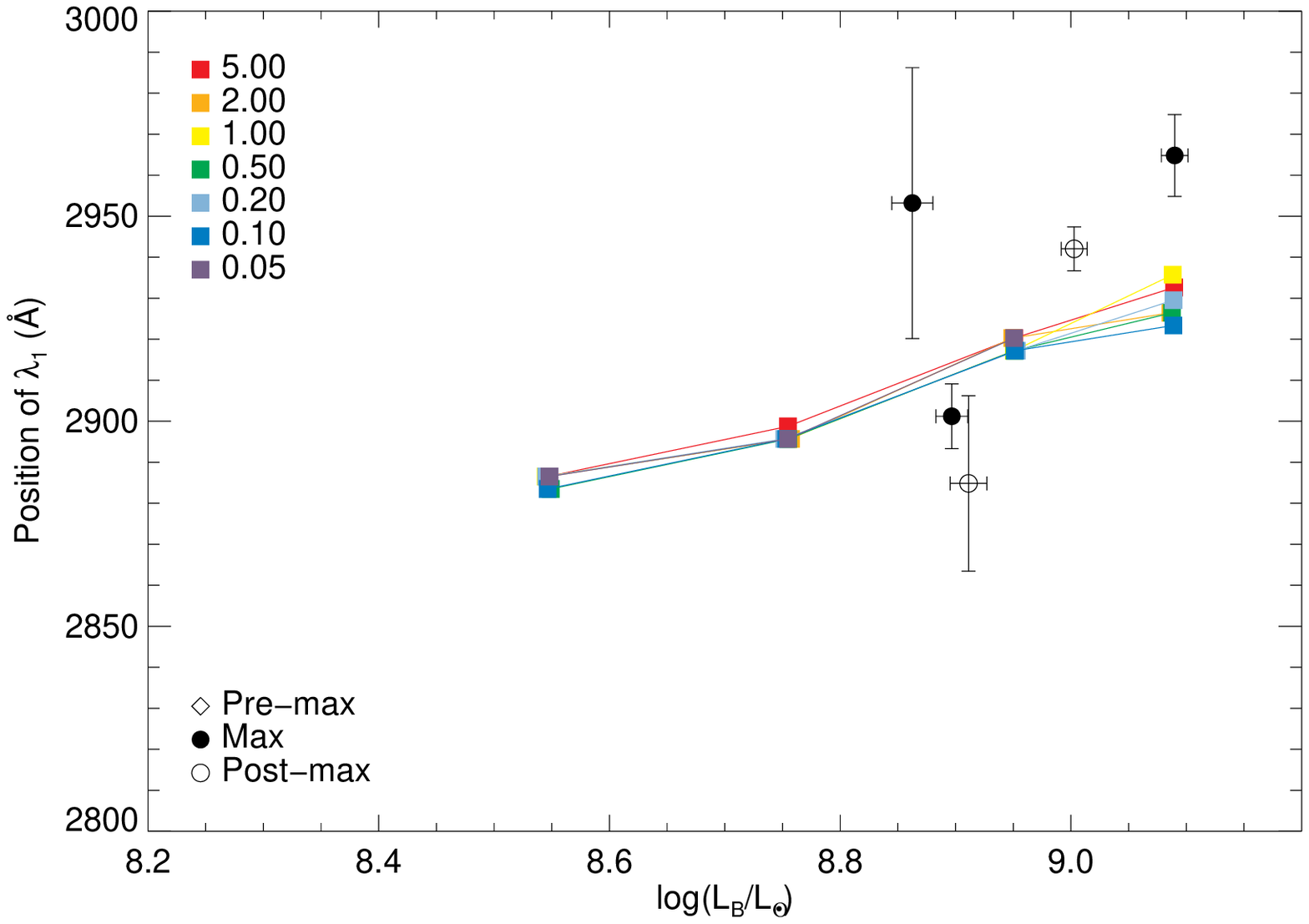}
    \includegraphics[width = \columnwidth]{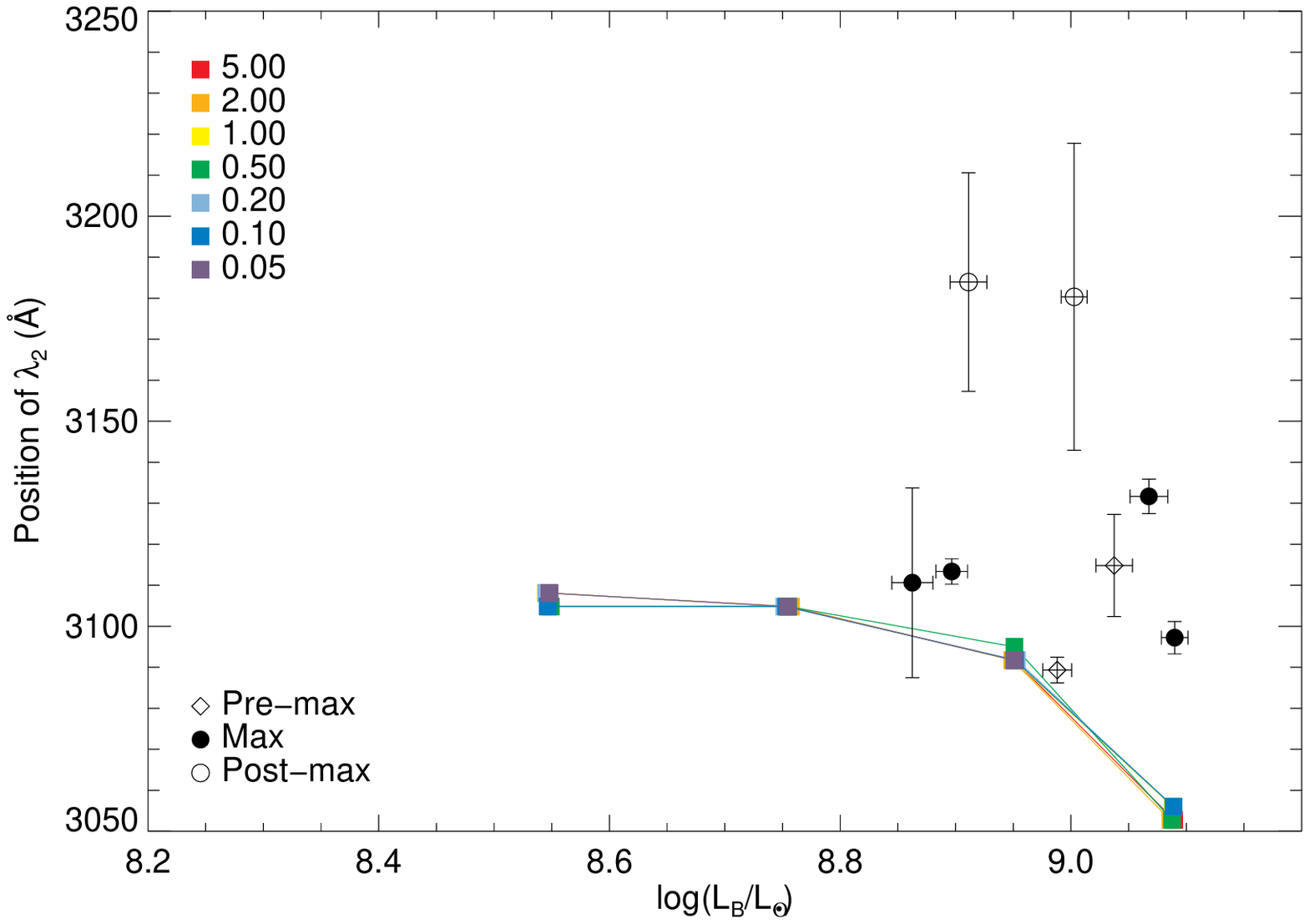}
 \includegraphics[width = \columnwidth]{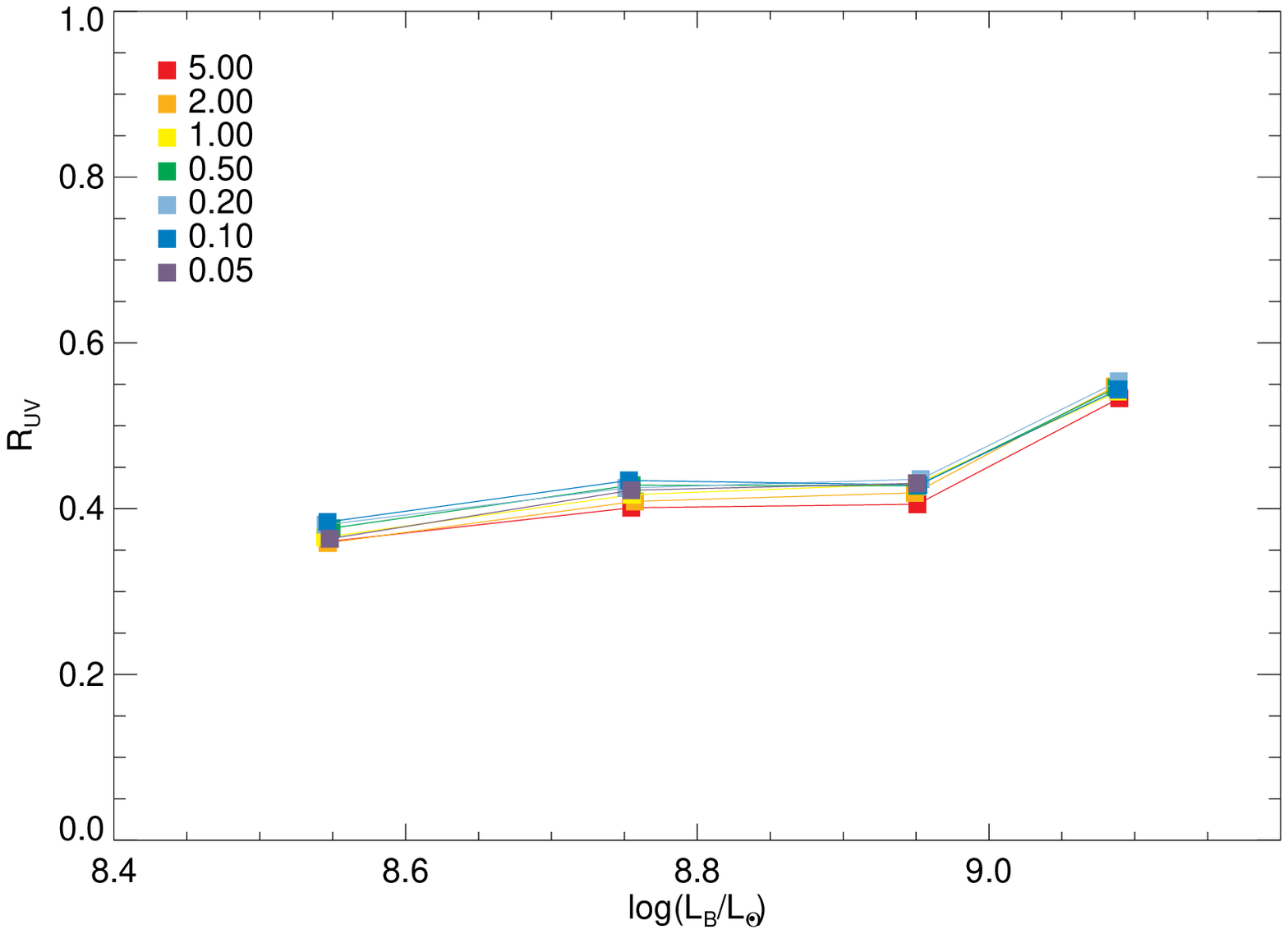}
  \caption{The effect of changing just the amount of manganese in the outer ejecta.  Upper -- UV colours; Middle -- The positions of $\lambda_1$ and $\lambda_2$;    Lower: $R_{UV}$ }
  \label{fig:Mn_seq}
\end{figure*}

\begin{figure*}
  \centering
  \includegraphics[width = \textwidth]{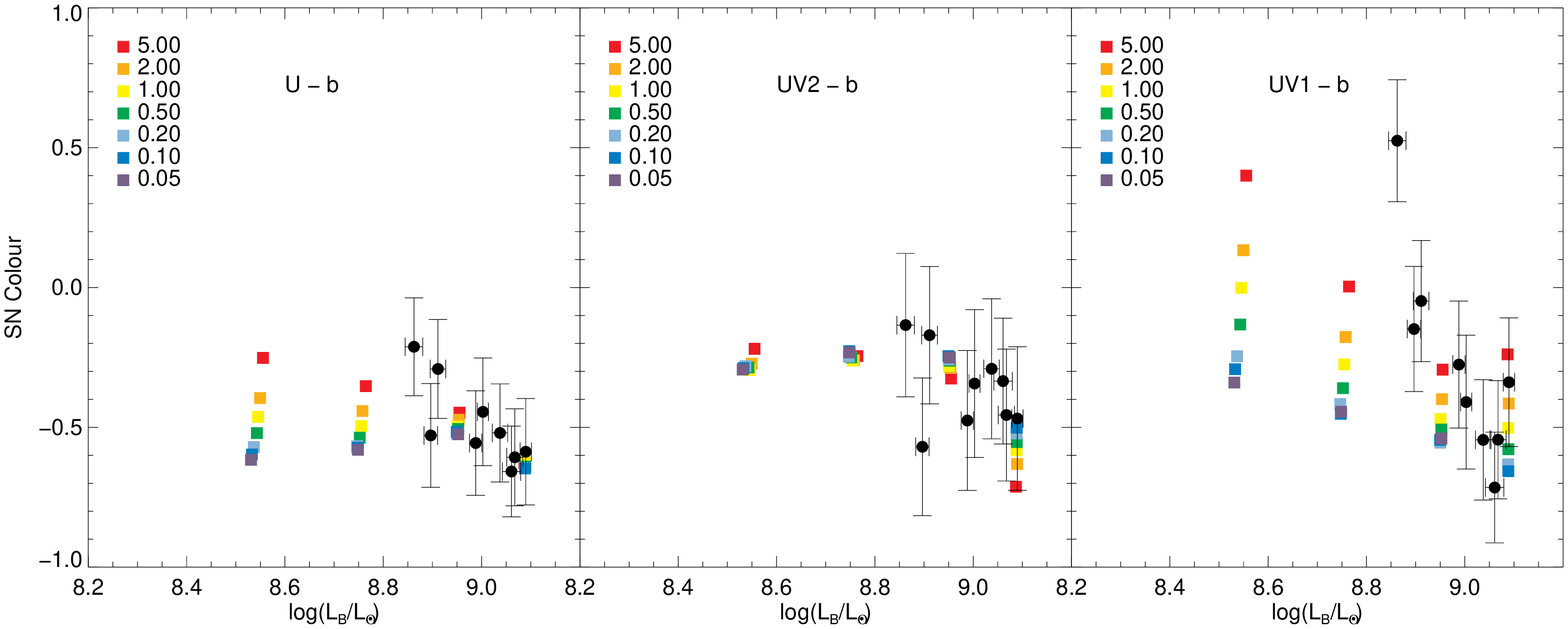}
  \includegraphics[width = \columnwidth]{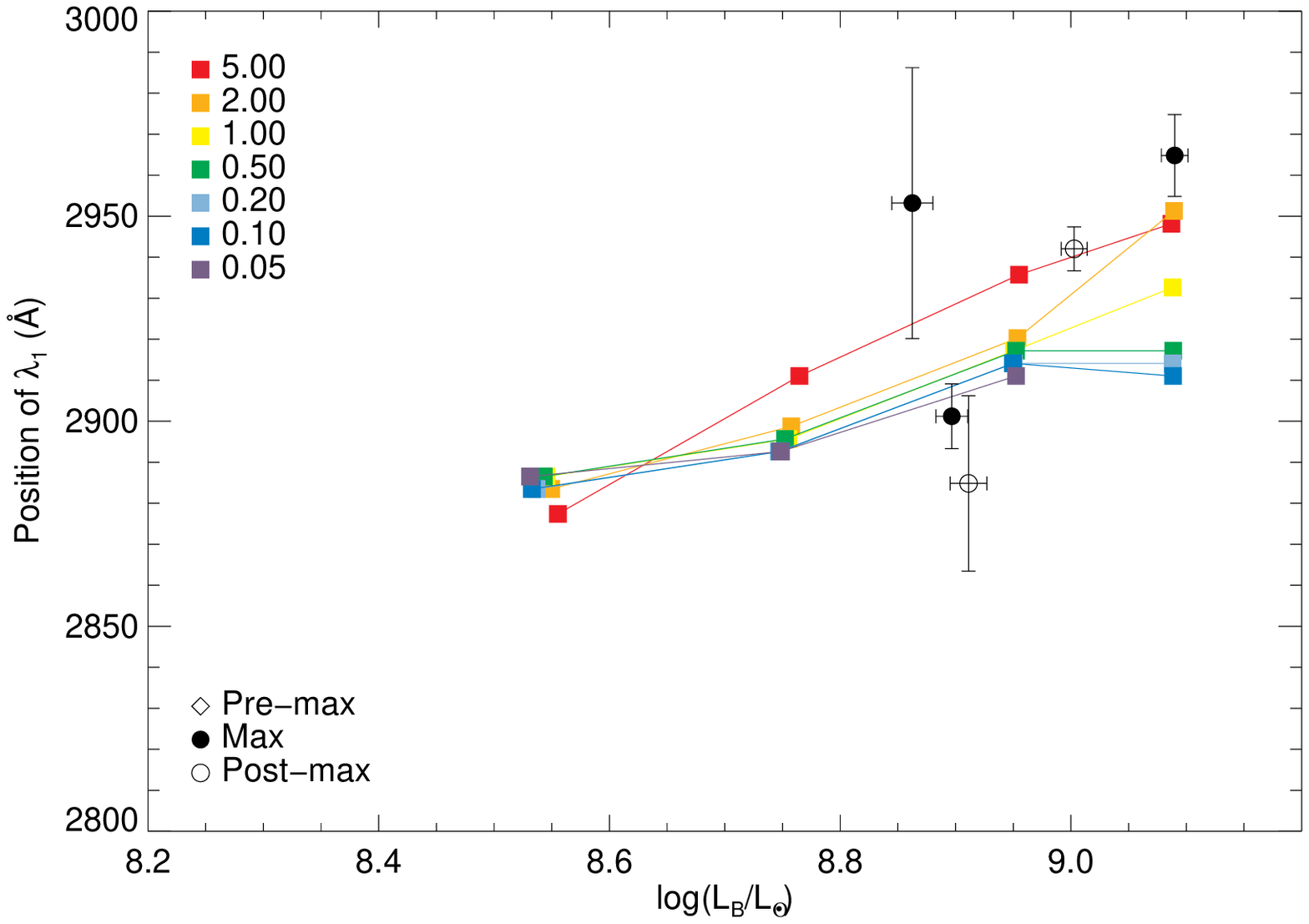}
    \includegraphics[width = \columnwidth]{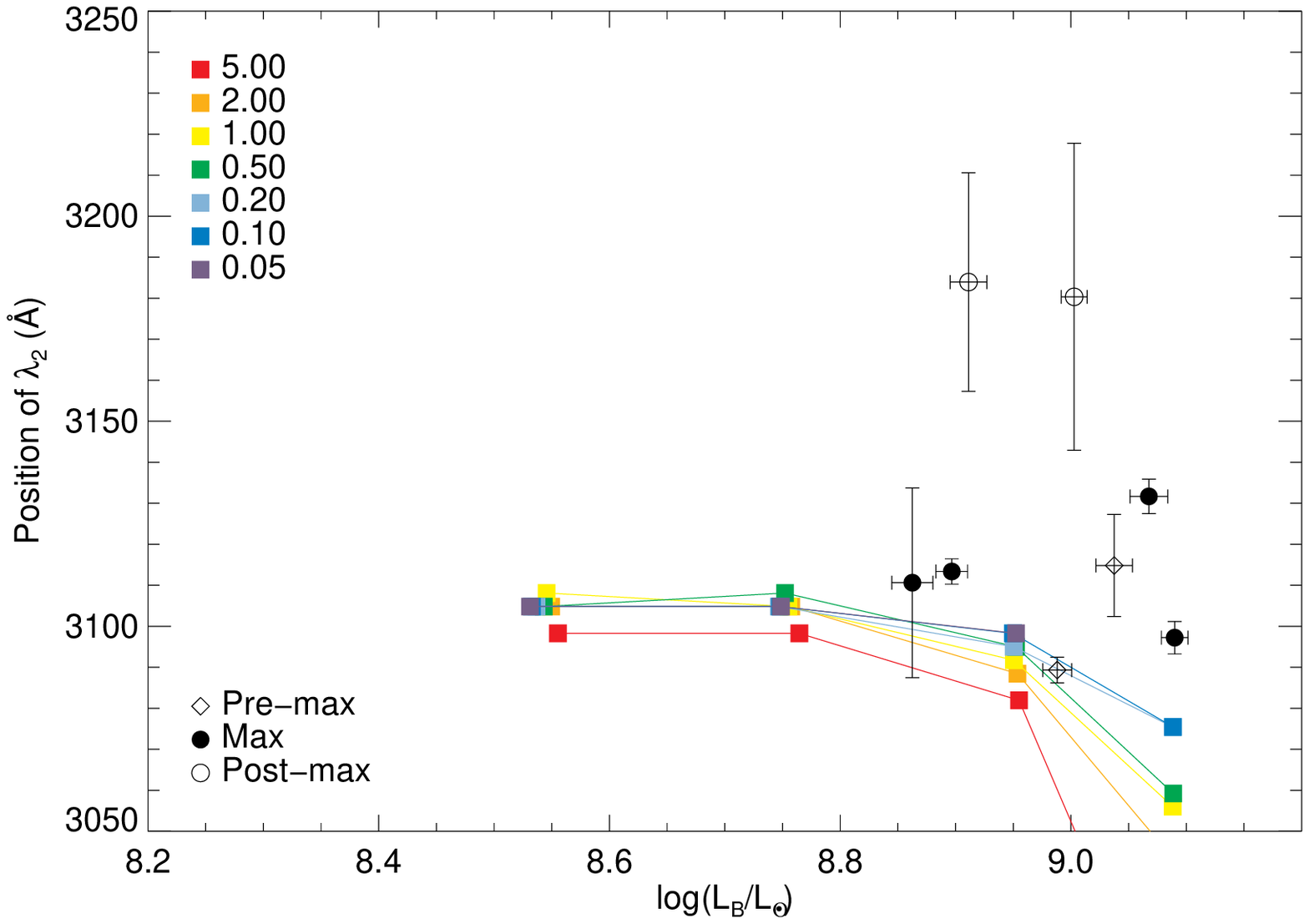}
 \includegraphics[width = \columnwidth]{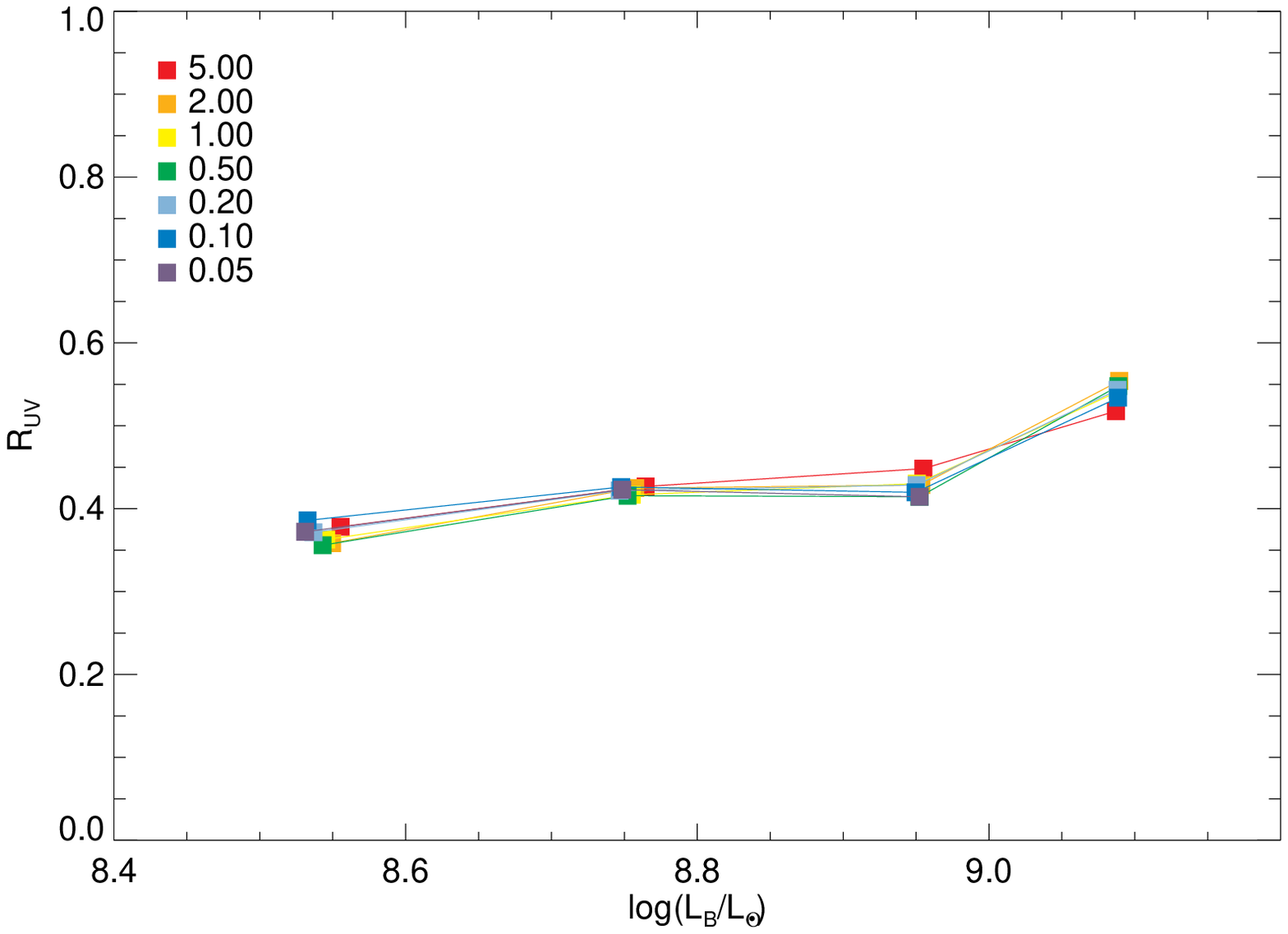}
  \caption{The effect of changing just the amount of \nickel\ in the outer ejecta.  Upper -- UV colours; Middle -- The positions of $\lambda_1$ and $\lambda_2$;    Lower: $R_{UV}$ }
  \label{fig:56Ni_seq}
\end{figure*}

\begin{figure*}
  \centering
  \includegraphics[width = \textwidth]{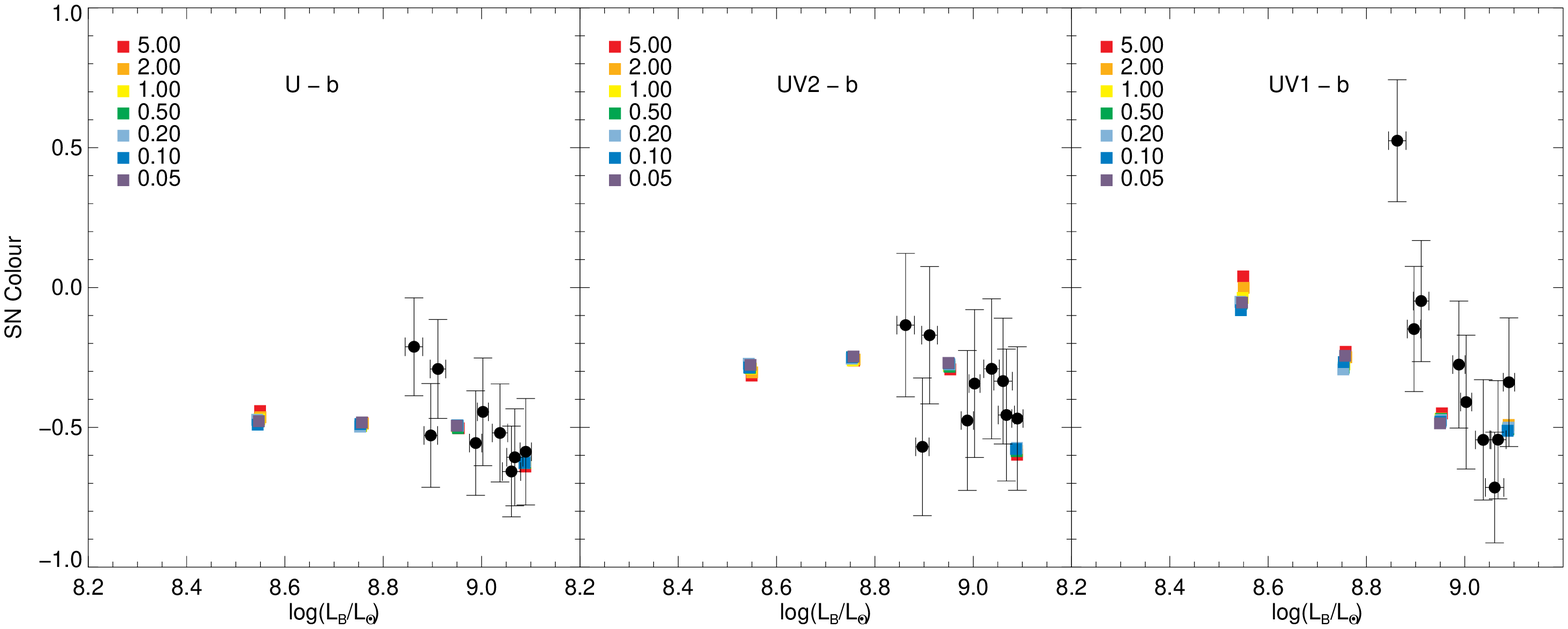}
  \includegraphics[width = \columnwidth]{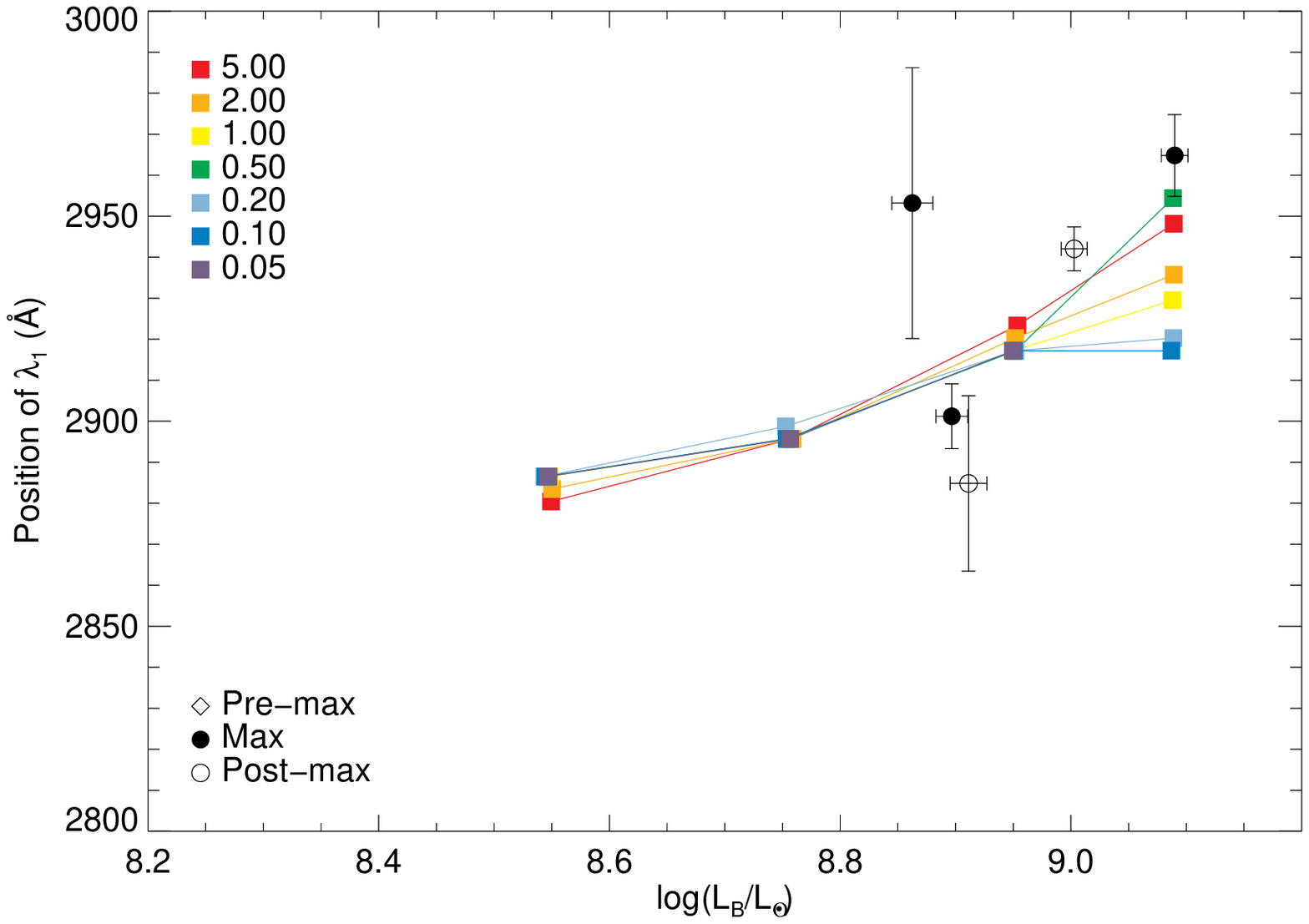}
    \includegraphics[width = \columnwidth]{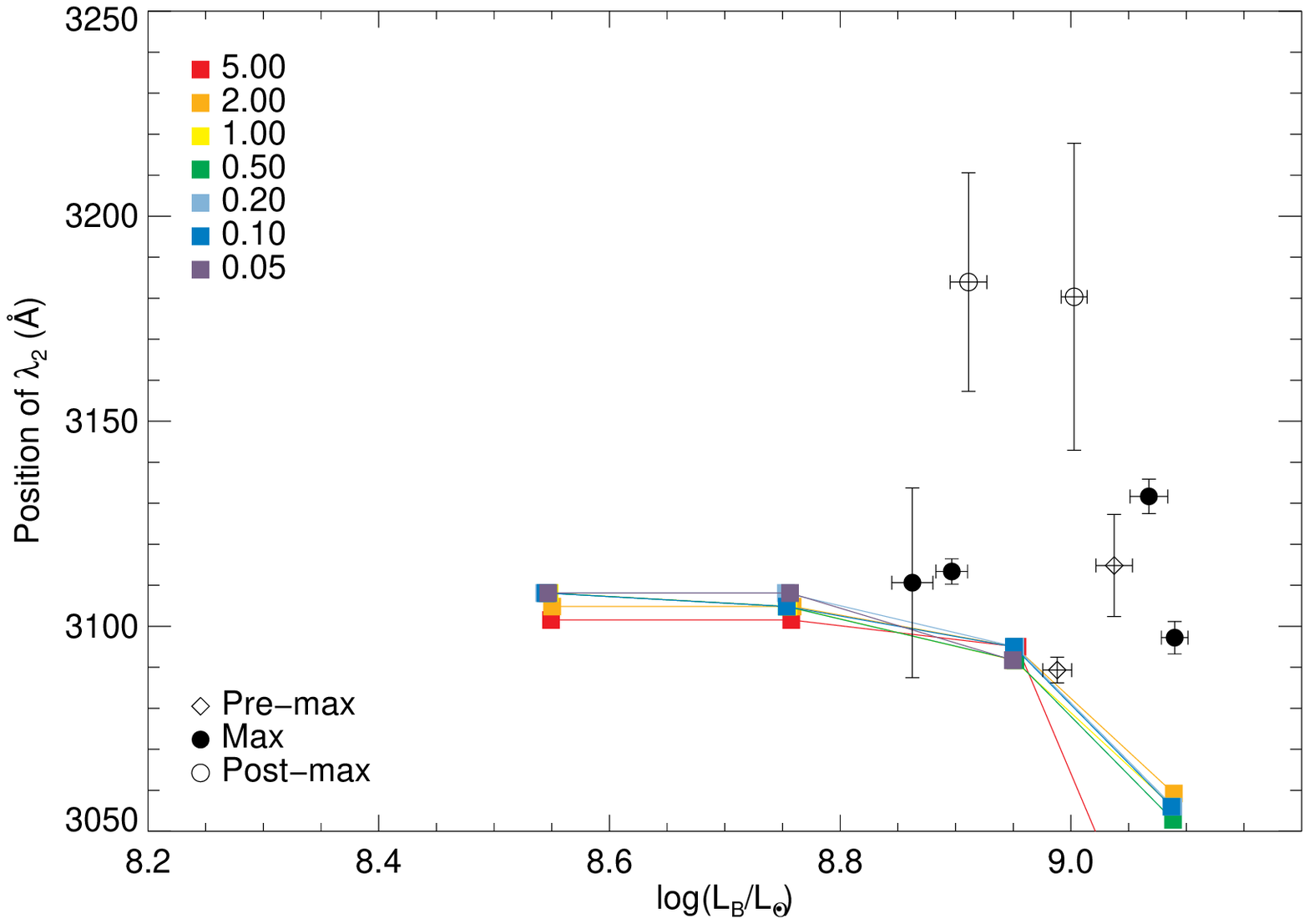}
 \includegraphics[width = \columnwidth]{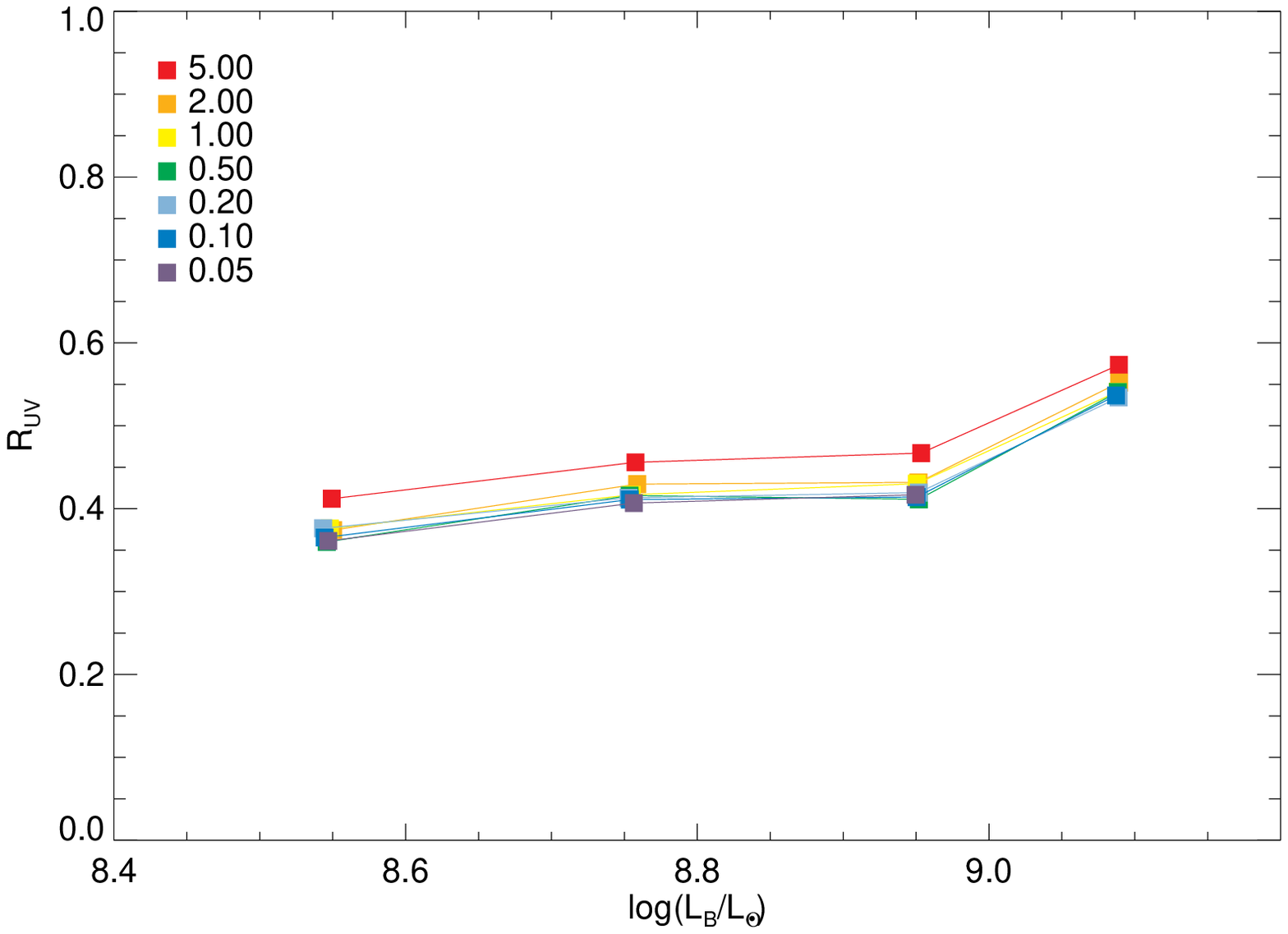}
  \caption{The effect of changing just the amount of titanium in the outer ejecta.  Upper -- UV colours; Middle -- The positions of $\lambda_1$ and $\lambda_2$;    Lower: $R_{UV}$ }
  \label{fig:Ti_seq}
\end{figure*}

Figures \ref{fig:cr_seq} -- \ref{fig:Ti_seq} show the effect that changing the amount of only 1 element has on the UV colours, positions of the UV emission features and the ratio $R_{UV}$.

\end{document}